\documentclass[aps,prc,showpacs,showkeys]{revtex4}

\usepackage{graphicx}
\usepackage{dcolumn}
\usepackage{amsthm}
\usepackage{bm}

\newtheorem*{Theorem}{Theorem}

\begin{document}

\title{Angular bremsstrahlung during $\alpha$-decay, emission from the internal region and unified formula of the bremsstrahlung probability}

\author{Sergei~P.~Maydanyuk\thanks{E-mail: maidan@kinr.kiev.ua}}


\affiliation{Institute for Nuclear Research, National Academy of Science of Ukraine} 


\date{\today}

\begin{abstract}
Model of angular bremsstrahlung of photons emitted during $\alpha$-decay is presented. A special emphasis is given on development of unified formalism of matrix elements in the dipole and multipolar approaches.
A probability of the emission of photons calculated on the basis of the multipole model without any normalization on experimental data (i.~e. in absolute scale) is found at $90^{\circ}$ of the angle $\vartheta_{\alpha\gamma}$ between directions of motion of the $\alpha$-particle (with its tunneling under barrier) and emission of photons to be in a good agreement with the newest experimental data for the $^{210}{\rm Po}$, $^{214}{\rm Po}$, and $^{226}{\rm Ra}$ nuclei. The spectrum for $^{244}{\rm Cm}$ is found at $\vartheta_{\alpha\gamma} = 25^{\circ}$ to be in satisfactory agreement with high limit of errors of experimental data of Japanese group.
A comparative analysis for the spectra calculated for $^{210}{\rm Po}$ by the multipole and dipole approaches in the absolute scale and with normalization on experimental data is performed.
The emission of photons from the internal well in the dipole approach is found to be not small while the multipolar approach does not show such a strong dependence.
Distribution of the bremsstrahlung probability on the numbers of protons and nucleons of the $\alpha$-decaying nucleus in selected region close to $^{210}{\rm Po}$ is obtained.
An unified formula of the bremsstrahlung probability during the $\alpha$-decay of arbitrary nucleus expressed directly through the $Q_{\alpha}$-value and numbers $A_{p}$, $Z_{p}$ of nucleons and protons of this nucleus
is proposed.
\end{abstract}



\maketitle


\section{Introduction
\label{introduction}}

Last two decades many experimental and theoretical efforts have been made in investigation of nature of bremsstrahlung emission which accompanies $\alpha$-decay of heavy nuclei (see references in~\cite{Maydanyuk.2006.EPJA,Maydanyuk.2009.NPA}). A key idea of such research consists in search of new information about dynamics of the $\alpha$-decay (and dynamics of tunneling) which is supposed to be extracted from analysis of the measured bremsstrahlung spectra. Tunneling times in nuclear collisions and decays are extremely small, close to nuclear time. This fact results in impossibility to test non-stationary methods of tunneling experimentally. However, researchers open new ways how to obtain new information about the dynamics of nuclear processes. An increasing interest in study of the bremsstrahlung processes accompanying the $\alpha$-decay could be explained by this idea mainly: to find a new approach how through analysis of the bremsstrahlung spectra to ``measure'' dynamics of the $\alpha$-decay (perhaps, in its first stage), to estimate duration of tunneling of the $\alpha$-particle through barrier.

Many approaches for description of the bremsstrahlung emission during the $\alpha$-decay have already been developed where models with semiclassical spherically symmetric description of the $\alpha$-decay are prevailing (see \cite{Dyakonov.1996.PRLTA,Dyakonov.1999.PHRVA,Bertulani.1999.PHRVA,Takigawa.1999.PHRVA}, also calculations of the spectra see in~\cite{Kasagi.1997.JPHGB}). In comparison with fully quantum approach, the semiclassical one allows to work with characteristics and parameters, physical sense of which is natural that simplifies this task, allowing to understand studied questions easier. Enough well description of experimental data has already been achieved in such approach, where one can note a resent success in agreement between theory and experiment for the controversial nucleus $^{210}{\rm Po}$ \cite{Boie.2007.PRL,Jentschura.2008.PRC}.
Perspectives are certain in study of dynamics of the $\alpha$-decay with some analysis of the bremsstrahlung~\cite{Bertulani.1999.PHRVA,Misicu.2001.JPHGB,Dijk.2003.FBSSE}, in study of dynamics of tunneling in the $\alpha$-decay~\cite{Serot.1994.NUPHA,Dijk.1999.PRLTA,Dijk.2002.PHRVA,Ivlev.2004.PHRVA},
in research of peculiarities of the polarized bremsstrahlung during $\alpha$-decay and influence of electron shells on it \cite{Amusia.2007.JETP}, in effect~\cite{Flambaum.1999.PRLTA} called as \emph{M\"{u}nchhausen effect} which increases penetrability of the barrier due to charged-particle emission during its tunneling and which could be interesting for further study of the photon bremsstrahlung during tunneling in the $\alpha$-decay.
However, the fully quantum approach (starting from \cite{Batkin.1986.SJNCA} and then \cite{Papenbrock.1998.PRLTA,Tkalya.1999.JETP}) seems to be the most accurate and motivated from the physical point of view in description of emission of photons, to be the richest in study of quantum properties and new effects. Among the fully quantum approaches a model proposed for the first time by Papenbrock and Bertsch in~\cite{Papenbrock.1998.PRLTA} has been developing the most intensively, where wave function of photons is used in the dipole approximation. In such dipole approach the matrix element is calculated with higher convergence and without visible decrease of accuracy, that makes this problem to be studied for many researchers in the fully quantum consideration.

If intensity of bremsstrahlung was varied enough visibly at change of the angle, then the emission of photons would take influence on dynamics of the $\alpha$-decay essentially and, therefore, change some its characteristic. From such point of view discussions~\cite{Eremin.2000.PRLTA,Kasagi.2000.PRLTA} show a way for obtaining the new information about the $\alpha$-decay: through \emph{angular analysis of the bremsstrahlung during the $\alpha$-decay}. But for such researches a model of description of the bremsstrahlung in the $\alpha$-decay which takes a value of the angle between the directions of the $\alpha$-particle propagation (or tunneling) and the photon emission into account, should be constructed.
In such direction two independent groups have developed the angular formalism in the semiclassical dipole approach including quadrupole term \cite{Boie.2007.PRL,Jentschura.2008.PRC} and in the fully quantum approach \cite{Maydanyuk.nucl-th.0404013,Maydanyuk.2006.EPJA,Maydanyuk.2008.EPJA,Maydanyuk.2008.MPLA,Maydanyuk.2009.NPA} where wave function of photons was mainly expanded by spherical waves.
In particular, the good agreement for the $^{210}{\rm Po}$ nucleus between experimental data and the calculated spectra in the approach \cite{Boie.2007.PRL,Jentschura.2008.PRC} was obtained, neglecting by emission of photons from the internal nuclear region before the barrier.
From such results it could follow that such internal emission of photons is extremely small and does not influence on the spectra. In such a case, it looks to be impossible to extract any useful information about processes of $\alpha$-decay inside the nuclear region before the barrier from experimental data of the bremsstrahlung.
But, in order to clarify this the emission of photons from internal region should be estimated on the basis of fully quantum calculations based on the realistic potential between $\alpha$-particle and daughter nucleus.
A question of influence of emission from the internal region of the total bremsstrahlung spectra on the basis of realistic $\alpha$-nucleus potential has not been studied yet.

Taking into account \emph{expansion in spherical waves} for description of angular correction of the wave function of photons~\cite{Maydanyuk.nucl-th.0404013} and realistic form of interaction between the $\alpha$-particle and the daughter nucleus~\cite{Maydanyuk.2006.EPJA}, we achieved a little better agreement between such fully quantum approach and later obtained experimental data~\cite{Boie.2007.PRL} for the $^{210}{\rm Po}$ nucleus (for explanation see Fig.~3 in \cite{Maydanyuk.2008.EPJA} and discussions here). Results in descriptions of the newest experimental data~\cite{Maydanyuk.2008.EPJA,Maydanyuk.2008.MPLA,Maydanyuk.2009.NPA} for the $^{214}{\rm Po}$ and $^{226}{\rm Ra}$ nuclei seem to be enough good also, where we have been achieving agreement between theory and experiment up to 765~keV. But, a \emph{multipole approach} (started from~\cite{Tkalya.1999.JETP}, then in~\cite{Maydanyuk.2003.PTP}) seems to be the most accurate and corrected in spatial description of the emission of photons during the $\alpha$-decay. Such approach turns out to calculate probability without any normalization relatively experimental data (i.~e. in co-called \emph{absolute scale}), and achieves enough good agreement with them.
This peculiarity adds power of prediction to the multipolar approach:
it allows to study the bremsstrahlung during the $\alpha$-decays of arbitrary nucleus,
to estimate the emission from the internal region,
to predict new spectra.

In~\cite{So_Kim.2000.JKPS} dependence of the bremsstrahlung probability on the electrical charge of the daughter nucleus was analyzed. But, it is unclear how much energy of the $\alpha$-particle takes influence on the photons emission. Calculating the bremsstrahlung probabilities for the different isotopes with different mass numbers, I have seen that this probability is determined by combination between numbers of protons and neutrons in the nucleus under decay, rather than by the electric charge of this nucleus, and one come to conclusion about direct dependence of the bremsstrahlung probability on the effective charge $Z_{\rm eff}$.
It could be interesting to clarify whether any other parameter or characteristic exists, influence of which on the probability is essential. More intriguing task has been appeared: \emph{to construct an unified formula of the bremsstrahlung probability during the $\alpha$-decay of the arbitrary nucleus which is directly expressed through all these parameters and characteristics}.
\emph{But, whether is it possible to describe the bremsstrahlung spectra for all different nuclei by only one formula in general? Whether is it possible to describe the bremsstrahlung spectrum for only one arbitrary nucleus with very high accuracy inside the energy region of the photon emitted used in experiments?}
To answer on such questions, it should be desirable to use the model which is the most accurate and corrected in description of this reaction.

This paper answers on these questions and it is organized so. In Sec.~II the model of the bremsstrahlung accompanying the $\alpha$-decay is presented, where emphasis is made on the angular formalism in framework of the dipole and
multipolar approaches of the matrix elements and calculation of the bremsstrahlung probability.
In Sec.~\ref{sec.3.3} the model is tested on experimental data \cite{Maydanyuk.2008.EPJA,Maydanyuk.2008.MPLA} for the $^{214}{\rm Po}$ and $^{226}{\rm Ra}$ nuclei.
The spectrum for $^{244}{\rm Cm}$ is added to such a picture, compared with high limit of errors of experimental data \cite{Kasagi.1997.JPHGB,Kasagi.1997.PRLTA}.
In Sec.~\ref{sec.3.4} distribution of the bremsstrahlung probability on the numbers of protons and nucleons of the $\alpha$-decaying nucleus in selected region close to $^{210}{\rm Po}$ is presented.
In Sec.~\ref{sec.3.5} the angular spectra for the $^{210}{\rm Po}$ nucleus by the multipolar model are presented.
In Sec.~\ref{sec.3.6} the formula of the bremsstrahlung probability during the $\alpha$-decay, based only on the $Q_{\alpha}$-value and numbers of protons and neutrons of the decaying nucleus, has been constructed. Inside region of the $\alpha$-active nuclei from $^{106}{\rm Te}$ up to the nucleus with numbers of nucleons and protons $A_{p}=266$ and $Z_{p}=109$ with energy of the photons emitted from 50 keV up to 900 keV good agreement has been achieved between the spectra, obtained on the basis of the multipole model, and the bremsstrahlung spectra obtained on the basis of the proposed formula.
At finishing, results are summarized.



\section{Model
\label{sec.2}}

\subsection{Matrix element of emission
\label{sec.2.1}}

We define probability of the bremsstrahlung emission during $\alpha$-decay of nucleus in terms of transition matrix element of the composite system ($\alpha$-particle and daughter nucleus) from its state before emission of photon (called as \emph{initial $i$-state}) into its state after such emission (called as \emph{final $f$-state}). In this paper, I shall use the definition of the matrix element like (2.11) in \cite{Maydanyuk.2003.PTP} (in the first correction of the non-stationary perturbation theory with stationary limits $t_{0}=-\infty$ and $t_{1}=+\infty$,
and with normalization $|C| \to 1$):
\begin{equation}
  a_{fi} = F_{fi} \cdot 2\pi \,\delta(w_{f}-w_{i}+w),
\label{eq.2.1.1}
\end{equation}
where
\begin{equation}
\begin{array}{lcl}
  F_{fi} & = &
    Z_{eff}\, \displaystyle\frac{e}{m} \,
    \sqrt{\displaystyle\frac{2\pi\hbar}{w}} \cdot p\,(k_{i},k_{f}), \\

  p\,(k_{i}, k_{f}) & = &
    \displaystyle\sum\limits_{\alpha=1,2} \mathbf{e}^{(\alpha),*}\, \mathbf{p}\,(k_{i}, k_{f}), \\

  \mathbf{p}\,(k_{i}, k_{f}) & = &
    \biggl< k_{f} \biggl| \,  e^{-i\mathbf{kr}} \displaystyle\frac{\partial}{\partial \mathbf{r}} \,
    \biggr| \,k_{i} \biggr> =
    \int
      \psi^{*}_{f}(\mathbf{r}) \:
      e^{-i\mathbf{kr}} \displaystyle\frac{\partial}{\partial \mathbf{r}}\:
      \psi_{i}(\mathbf{r}) \;
      \mathbf{dr}
\end{array}
\label{eq.2.1.2}
\end{equation}
and $\psi_{i}(\mathbf{r}) = |k_{i}\bigr>$ and $\psi_{f}(\mathbf{r}) = |k_{f}\bigr>$ are stationary wave functions of the decaying system in the initial $i$-state and final $f$-state which do not contain number of photons emitted,
$Z_{\rm eff}$ and $m$ are effective charge and reduced mass of this system.
$\mathbf{e}^{(\alpha)}$ are unit vectors of polarization of the photon emitted, $\mathbf{k}$ is wave vector of the photon and $w = k = \bigl| \mathbf{k}\bigr|$. Vectors $\mathbf{e}^{(\alpha)}$ are perpendicular to $\mathbf{k}$ in Coulomb calibration. We have two independent polarizations $\mathbf{e}^{(1)}$ and $\mathbf{e}^{(2)}$ for the photon with impulse $\mathbf{k}$ ($\alpha=1,2$).
One can develop formalism simpler in the system of units where $\hbar = 1$ and $c = 1$, but we shall write constants $\hbar$ and $c$ explicitly.
Let's find also square of the matrix element $a_{fi}$ used in definition of \emph{probability of transition}.
Using the \emph{formula of power reduction of $\delta$-function} (see~\cite{Bogoliubov.1980}, \S~21, p.~169):
\begin{equation}
  [\delta(w)]^{2} = \delta(w)\: \delta(0) = \delta(w) \: (2\pi)^{-1} \int dt =
  \delta(w)\: (2\pi)^{-1}\, T,
\label{eq.2.1.3}
\end{equation}
we find ($T\to +\infty$ is higher time limit):
\begin{equation}
  |a_{fi}|^{2} = 2\pi\: T\: |F_{fi}|^{2} \cdot \delta(w_{f}-w_{i}+w),
\label{eq.2.1.4}
\end{equation}
that looks like (4.21) in \cite{Bogoliubov.1980} (with accuracy up to factor $(2\pi)^{2}$) and
like (42.5) in \cite{Landau.v3.1989} (exactly, see \S~42, p.~189).


\subsection{Linear and circular polarizations of the photon emitted
\label{sec.2.2}}

Rewrite vectors of \emph{linear} polarization $\mathbf{e}^{(\alpha)}$ through \emph{vectors of circular polarization} $\mathbf{\xi}_{\mu}$ with opposite directions of rotation (see ref.~\cite{Eisenberg.1973}, (2.39), p.~42):
\begin{equation}
\begin{array}{ccc}
  \mathbf{\xi}_{-1} = \displaystyle\frac{1}{\sqrt{2}}\,
                      \bigl(\mathbf{e}^{(1)} - i\mathbf{e}^{(2)}\bigr), &
  \mathbf{\xi}_{+1} = -\displaystyle\frac{1}{\sqrt{2}}\,
                      \bigl(\mathbf{e}^{(1)} + i\mathbf{e}^{(2)}\bigr), &
  \mathbf{\xi}_{0} = \mathbf{e}^{(3)} = 0.
\end{array}
\label{eq.2.2.1}
\end{equation}
Then $p\,(k_{i},k_{f})$ can be rewritten so:
\begin{equation}
  p\,(k_{i}, k_{f}) =
    \sum\limits_{\mu = -1, 1}  h_{\mu}\,\mathbf{\xi}^{*}_{\mu}
    \int
      \psi^{*}_{f}(\mathbf{r})\:
      e^{-i\mathbf{kr}} \displaystyle\frac{\partial}{\partial \mathbf{r}} \:
      \psi_{i}(\mathbf{r}) \;
    \mathbf{dr},
\label{eq.2.2.2}
\end{equation}
\begin{equation}
\begin{array}{ccc}
  h_{\pm} = \mp \displaystyle\frac{1 \pm i}{\sqrt{2}}, &
  h_{-1} + h_{+1} = -i\sqrt{2}, &
  \sum\limits_{\alpha = 1,2} \mathbf{e}^{(\alpha),*} =
    h_{-1} \mathbf{\xi}_{-1}^{*} + h_{+1} \mathbf{\xi}_{+1}^{*}.
\end{array}
\label{eq.2.2.3}
\end{equation}


\subsection{Different expansions of the vector potential $\mathbf{A}$
\label{sec.2.3}}

In order to simplify calculations of the matrix element $p\,(k_{i}, k_{f})$ with taking into account angular correlation between vectors $\mathbf{k}$ and $\mathbf{r}$, function $e^{-i\mathbf{kr}}$ connected with the vector potential $\mathbf{A}$ of the electro-magnetic field of the daughter nucleus should be expanded into basis of functions.
At present, only three different types of expansion have been used in this problem:
\begin{itemize}
\item
The expansion in the dipole approximation:
\begin{equation}
  e^{i \mathbf{kr}} = 1 + i\,\mathbf{kr} + \ldots
\label{eq.2.3.1}
\end{equation}
where the first item is found to be mainly used correctly in further calculations of the matrix element;

\item
The expansion by the spherical waves (in direction of papers~\cite{Maydanyuk.nucl-th.0404013,Maydanyuk.2006.EPJA,Maydanyuk.2008.EPJA,Maydanyuk.2008.MPLA,Maydanyuk.2009.NPA}):
\begin{equation}
  e^{i \mathbf{kr}} =
    \displaystyle\sum_{l=0}^{+\infty}\,
    i^{l} (2l+1)\, P_{l}(\cos \theta_{\alpha\gamma}) \, j_{l}(kr),
\label{eq.2.3.2}
\end{equation}
where $\theta_{\alpha\gamma}$ is angle between vectors $\mathbf{k}$ and $\mathbf{r}$
($\mathbf{kr} = kr\,\cos{\theta_{\alpha\gamma}}$);

\item
The expansion by electric and magnetic multiples
(see ref.\cite{Eisenberg.1973}, (2.106) in p.~58):
\begin{equation}
  \mathbf{\xi}_{\mu}\, e^{i \mathbf{kr}} =
    \mu\, \sqrt{2\pi}\, \sum_{l, \nu}\,
    (2l+1)^{1/2}\, i^{l}\,  D_{\nu\mu}^{l} (\varphi,\theta,0) \cdot
    \Bigl[ \mathbf{A}_{l\nu} (\mathbf{r}, M) +
    i\mu\, \mathbf{A}_{l\nu} (\mathbf{r}, E) \Bigr],
\label{eq.2.3.3}
\end{equation}
where (see ref.\cite{Eisenberg.1973}, (2.73) in p.~49, (2.80) in p.~51)
\begin{equation}
\begin{array}{lcl}
  \mathbf{A}_{l\nu}(\mathbf{r}, M) & = &
        j_{l}(kr) \: \mathbf{T}_{ll,\nu} ({\mathbf n}_{ph}), \\
  \mathbf{A}_{l\nu}(\mathbf{r}, E) & = &
        \sqrt{\displaystyle\frac{l+1}{2l+1}}\,
        j_{l-1}(kr) \: \mathbf{T}_{ll-1,\nu}({\mathbf n}_{ph}) -
        \sqrt{\displaystyle\frac{l}{2l+1}}\,
        j_{l+1}(kr) \: \mathbf{T}_{ll+1,\nu}({\mathbf n}_{ph}).
\end{array}
\label{eq.2.3.4}
\end{equation}
\end{itemize}
Here, $\mathbf{A}_{l\nu}(\textbf{r}, M)$ and $\mathbf{A}_{l\nu}(\textbf{r}, E)$ are \emph{magnetic} and \emph{electric multipoles}, $j_{l}(kr)$ are \emph{spherical Bessel functions of order $l$}, $\mathbf{T}_{ll',\nu}(\mathbf{n})$ are \emph{vector spherical harmonics},
$\theta_{1}$, $\theta_{2}$, $\theta_{3}$ are angles defining direction of vector $\mathbf{k}$ relatively axis $z$ in selected frame system.

In this paper, I shall use the multipole expansion assuming that it gives us the most reliable approach to describe the angular emission of photon during the $\alpha$-decay.
Matrix-function $D_{\nu\mu}^{l,*}(\varphi,\theta,0)$ defines direction of vector $\mathbf{k}$ relatively axis $z$ in the frame system for $\mathbf{r}$: angles $\varphi$ and $ \theta$ point to direction of vector $\mathbf{k}$, but not the vector $\mathbf{r}$.
The functions $\mathbf{T}_{ll',\nu}(\mathbf{n})$ have the following form (${\mathbf \xi}_{0} = 0$, see ref.\cite{Eisenberg.1973}, p.~45):
\begin{equation}
  \mathbf{T}_{jl,m} (\mathbf{n}) =
  \sum\limits_{\mu = \pm 1} (l, 1, j \,\big| \,m-\mu, \mu, m) \; Y_{l,m-\mu}(\mathbf{n}) \; \mathbf{\xi}_{\mu},
\label{eq.2.3.5}
\end{equation}
where $(l, 1, j \,\bigl| \, m-\mu, \mu, m)$ are \emph{Clebsh-Gordon coefficients} and
$Y_{lm}(\theta, \varphi)$ are \emph{spherical functions} defined, according to~\cite{Landau.v3.1989} (see p.~119, (28,7)--(28,8)).


\subsection{Approximation of the spherically symmetric $\alpha$-decay
\label{sec.2.4}}

Now we shall study the $\alpha$-decay in the spherically symmetric approximation.
We orientate the frame system so that axis $z$ will be parallel to the vector $\mathbf{k}$ and
\begin{equation}
  D_{\nu\mu}^{l}(\varphi,\theta,0) = \delta_{\mu\nu}.
\label{eq.2.4.1}
\end{equation}
In the spherically symmetric approximation, wave functions of the decaying system in the initial and final states are separated into the radial and angular components, and these states are characterized by quantum numbers $l$ and $m$. We shall be interesting in such photon emission when the system transits to superposition of all possible final states with different values of the magnetic numbers $m_{f}$ at the same orbital number $l_{f}$. Let's assume that in the initial state we have $l_{i}=m_{i}=0$ and the radial component of wave function $\varphi_{f} (r)$ does not depend on $m_{f}$ for selected $l_{f}$.
We write wave functions so:
\begin{equation}
\begin{array}{lcl}
  \psi_{i} (\mathbf{r}) & = & \varphi_{i} (r) \: Y_{00}({\mathbf n}_{r}^{i}), \\
  \psi_{f} (\mathbf{r}) & = & \varphi_{f} (r) \: \displaystyle\sum\limits_{m} Y_{l_{f}m}({\mathbf n}_{r}^{f})
\end{array}
\label{eq.2.4.2}
\end{equation}
and obtain:
\begin{equation}
  p\,(k_{i}, k_{f}) = \sqrt{2\pi}\: \sum\limits_{l} \: (-i)^{l}\, \sqrt{2l+1} \: \Bigl[ p_{l}^{M} -ip_{l}^{E} \Bigr],
\label{eq.2.4.3}
\end{equation}
where
\begin{equation}
\begin{array}{ll}
  p_{l}^{M} = \sum\limits_{\mu = -1, 1} \mu\, h_{\mu}\: p_{l\mu}^{M}, &
  p_{l}^{E} = \sum\limits_{\mu = -1, 1} \mu^{2} h_{\mu}\: p_{l\mu}^{E}
\end{array}
\label{eq.2.4.4}
\end{equation}
and
\begin{equation}
\begin{array}{lcl}
  p_{l\mu}^{M} & = &
        \displaystyle\int\limits^{+\infty}_{0} dr
        \displaystyle\int d\Omega \: r^{2} \,
        \psi^{*}_{f}(\mathbf{r}) \,
        \biggl( \displaystyle\frac{\partial}{\partial \mathbf{r}}\, \psi_{i}(\mathbf{r}) \biggr) \,
        \mathbf{A}_{l\mu}^{*} (\mathbf{r}, M), \\

  p_{l\mu}^{E} & = &
        \displaystyle\int\limits^{+\infty}_{0} dr
        \displaystyle\int d\Omega \: r^{2} \,
        \psi^{*}_{f}(\mathbf{r}) \,
        \biggl( \displaystyle\frac{\partial}{\partial \mathbf{r}}\, \psi_{i}(\mathbf{r}) \biggr)\,
        \mathbf{A}_{l\mu}^{*} (\mathbf{r}, E).
\end{array}
\label{eq.2.4.5}
\end{equation}

Using \emph{gradient formula} (see (2.56), p.~46 in~\cite{Eisenberg.1973}):
\begin{equation}
  \displaystyle\frac{\partial}{\partial \mathbf{r}}\:
    f(r)\, Y_{lm}({\mathbf n}_{r}) =
  \sqrt{\displaystyle\frac{l}{2l+1}}\:
    \biggl( \displaystyle\frac{df}{dr} + \displaystyle\frac{l+1}{r} f \biggr)\,
    \mathbf{T}_{l l-1, m}({\mathbf n}_{r}) -
  \sqrt{\displaystyle\frac{l+1}{2l+1}}\:
    \biggl( \displaystyle\frac{df}{dr} - \displaystyle\frac{l}{r} f \biggr)\,
    \mathbf{T}_{l l+1, m}({\mathbf n}_{r}),
\label{eq.2.4.6}
\end{equation}
we obtain:
\begin{equation}
  \displaystyle\frac{\partial}{\partial \mathbf{r}}\: \psi_{i}(\mathbf{r}) =
  -\,\displaystyle\frac{d\,\varphi_{i}(r)}{dr}\: \mathbf{T}_{01,0}(\mathbf{n}^{i}_{r}),
\label{eq.2.4.7}
\end{equation}
and then we calculate the matrix components:
\begin{equation}
\begin{array}{lcl}
  p_{l_{ph}}^{M} & = & -\; I_{M}(l_{f},l_{ph}, l_{ph}) \cdot J(l,l), \\
  p_{l_{ph}}^{E} & = &
    -\sqrt{\displaystyle\frac{l_{ph}+1}{2l_{ph}+1}}\; I_{E}(l_{f},l_{ph},l_{ph}-1) \cdot J(l_{f},l_{ph}-1)\; +
    \sqrt{\displaystyle\frac{l_{ph}}{2l_{ph}+1}}\; I_{E}(l_{f},l_{ph},l_{ph}+1) \cdot J(l_{f},l_{ph}+1),
\end{array}
\label{eq.2.4.8}
\end{equation}
where
\begin{equation}
\begin{array}{lcl}
  J(l_{f},n) & = &
  \displaystyle\int\limits^{+\infty}_{0}
    \varphi^{*}_{f}(l,r)\, \displaystyle\frac{d\varphi_{i}(r)}{dr}\:
    j_{n}(kr)\; r^{2} dr, \\

  I_{M}(l_{f}, l_{ph}, n) & = &
    \displaystyle\sum\limits_{\mu = \pm 1}
    \mu\, h_{\mu}
    \displaystyle\int
    Y_{l_{f}m}^{*}({\mathbf n}_{r}^{f}) \:
    \mathbf{T}_{01,0}(\mathbf{n}^{i}_{r})
    \: \mathbf{T}_{l_{ph} n,\mu}^{*}({\mathbf n}_{ph}) \: d\Omega, \\

  I_{E}(l_{f}, l_{ph}, n) & = &
    \displaystyle\sum\limits_{\mu = \pm 1}
    h_{\mu}
    \displaystyle\int
    Y_{l_{f}m}^{*}({\mathbf n}_{r}^{f}) \:
    \mathbf{T}_{01,0}(\mathbf{n}^{i}_{r})
    \: \mathbf{T}_{l_{ph} n,\mu}^{*}({\mathbf n}_{ph}) \: d\Omega.
\end{array}
\label{eq.2.4.9}
\end{equation}

Using the following value of the Clebsh-Gordon coefficient (see Appendix~A): 
\begin{equation}
  (110\,|1, -1, 0) = (110\,|-1, 1, 0) = \sqrt{\displaystyle\frac{1}{3}},
\label{eq.2.4.10}
\end{equation}
from (\ref{eq.2.3.5}) and (\ref{eq.2.4.7}) we obtain:
\begin{equation}
\begin{array}{c}
  \mathbf{T}_{01,0}(\mathbf{n}^{i}_{r}) =
    \displaystyle\sum\limits_{\mu = \pm 1} (110\,|-\mu\mu 0) \:
      Y_{1,-\mu}(\mathbf{n}^{i}_{r}) \: \mathbf{\xi}_{\mu} =
    \sqrt{\displaystyle\frac{1}{3}}
      \displaystyle\sum\limits_{\mu = \pm 1}
      Y_{1,-\mu}(\mathbf{n}^{i}_{r}) \: \mathbf{\xi}_{\mu}, \\

  \displaystyle\frac{\partial}{\partial \mathbf{r}}
    \psi_{i}(\mathbf{r}) =
  -\,\sqrt{\displaystyle\frac{1}{3}}\:
    \displaystyle\frac{d\varphi_{i}(r)}{dr}
    \displaystyle\sum\limits_{\mu = -1, 1}
    Y_{1,-\mu}(\mathbf{n}^{i}_{r}) \: \mathbf{\xi}_{\mu}
\end{array}
\label{eq.2.4.11}
\end{equation}
and for the angular integrals for transition into the superposition of all possible final $f$-states with different $m_{f}$ at the same $l_{f}$ from eq.~(\ref{eq.2.4.9}) we obtain:
\begin{equation}
\begin{array}{lcl}
  I_{M}(l_{f},l_{ph},n) & = &
    \sqrt{\displaystyle\frac{1}{3}}
    \displaystyle\sum\limits_{\mu = \pm 1} \mu\, h_{\mu}
    \sum\limits_{\mu^{\prime} = \pm 1} (n, 1, l_{ph} \big|\, \mu-\mu^{\prime}, \mu^{\prime}, \mu) \;
    \displaystyle\int \:
      Y_{l_{f}m}^{*}({\mathbf n}_{r}^{f}) \,
      Y_{1,-\mu^{\prime}}(\mathbf{n}_{r}^{i}) \,
      Y_{n, \mu-\mu^{\prime}}^{*}(\mathbf{n}_{ph}) \;
      d\Omega, \\

  I_{E}(l_{f},l_{ph},n) & = &
    \sqrt{\displaystyle\frac{1}{3}}
    \displaystyle\sum\limits_{\mu = \pm 1} h_{\mu}
    \sum\limits_{\mu^{\prime} = \pm 1} (n, 1, l_{ph} \big|\, \mu-\mu^{\prime}, \mu^{\prime}, \mu) \;
    \displaystyle\int \:
      Y_{l_{f}m}^{*}({\mathbf n}_{r}^{f}) \,
      Y_{1,-\mu^{\prime}}(\mathbf{n}_{r}^{i}) \,
      Y_{n, \mu-\mu^{\prime}}^{*}(\mathbf{n}_{ph}) \;
      d\Omega.
\end{array}
\label{eq.2.4.12}
\end{equation}

\subsection{Calculations of the angular integrals
\label{sec.2.5}}

Let us analyze a physical sense of vectors $\mathbf{n}^{i}_{r}$, $\mathbf{n}^{f}_{r}$ and $\mathbf{n}_{ph}$. According to definition of wave functions $\psi_{i} (\mathbf{r})$ and $\psi_{f} (\mathbf{r})$, the vectors $\mathbf{n}^{i}_{r}$ and $\mathbf{n}^{f}_{r}$ determine orientation of radius-vector $\mathbf{r}$ from the center of frame system to point where this wave functions describes the particle before and after the emission of photon. Such description of the particle has a probabilistic sense and is fulfilled over whole space.
Change of direction of motion (or tunneling) of the particle in result of the photon emission can be characterized by change of quantum numbers $l$ and $m$ in the angular wave function: $Y_{00}(\mathbf{n}^{i}_{r}) \to Y_{lm}(\mathbf{n}^{f}_{r})$ (which changes the probability of appearance of this particle along different directions, and angular asymmetry is appeared).
The vector $\mathbf{n}_{ph}$ determines orientation of radius-vector $\mathbf{r}$ from the center of the frame system to point where wave function of photon describes its ``appearance''. Using such a logic, we have:
\begin{equation}
  \mathbf{n}_{ph} = \mathbf{n}^{i}_{r} = \mathbf{n}^{f}_{r} = \mathbf{n}_{r}.
\label{eq.2.5.1}
\end{equation}
As we use the frame system where axis $z$ is parallel to vector $\mathbf{k}$ of the photon emission, then dependent on $\mathbf{r}$ integrant function in the matrix element represents amplitude (its square is probability) of appearance of the particle at point $\mathbf{r}$ after emission of photon, if this photon has emitted along axis $z$. Then angle $\theta$ (of vector $\mathbf{n}_{\mathbf{r}}$) is the angle between direction of the particle motion (with possible tunneling) and direction of the photon emission.

Let us consider the angular integral in (\ref{eq.2.4.12}) over $d\,\Omega$. Using (\ref{eq.2.5.1}), we find:
\begin{equation}
\begin{array}{l}
  \displaystyle\int \:
    Y_{lm}^{*}({\mathbf n}_{r}) \,
    Y_{1,-\mu^{\prime}}(\mathbf{n}_{r}) \,
    Y_{n, \mu-\mu^{\prime}}^{*}(\mathbf{n_{r}}) \;
    d\Omega = \\

  = (-1)^{l+n-\mu^{\prime}+1 + \frac{|m+\mu^{\prime}|}{2}} \; i^{l+n+1} \;
    \sqrt{\displaystyle\frac{3\,(2l+1)\,(2n+1)}{32\pi}\;
          \displaystyle\frac{(l-1)!}{(l+1)!} \;
          \displaystyle\frac{(n-|m+\mu^{\prime}|)!}{(n+|m+\mu^{\prime}|)!}} \;
    \times \\
  \;\times
    \displaystyle\int\:
      P_{l}^{1}(\cos{\theta}) \; P_{1}^{1}(\cos{\theta}) \; P_{n}^{|m+\mu^{\prime}|} (\cos{\theta}) \cdot
      \sin{\theta} \, d\theta \,d\varphi,
\end{array}
\label{eq.2.5.2}
\end{equation}
where $P_{l}^{m}(\cos{\theta})$ are \emph{associated Legandre's polynomial} (see \cite{Landau.v3.1989}, p.~752--754, (c,1)--(c,4); also see~\cite{Eisenberg.1973} (2.6), p.~34) and the following restrictions on possible values of $m$ and $l_{f}$
have been obtained:
\begin{equation}
\begin{array}{ccc}
  m = -\mu = \pm 1,  &  l_{f} \ge 1, &
  n \ge |\mu - \mu^{\prime}| = |m + \mu^{\prime}|.
\end{array}
\label{eq.2.5.3}
\end{equation}

Let's introduce the following differential matrix elements $dp_{l}^{M}$ and $dp_{l}^{E}$ dependent on the
angle $\theta$ :
\begin{equation}
\begin{array}{lcl}
  \displaystyle\frac{d \,p_{l}^{M}}{\sin{\theta}\,d\theta} & = &
    i^{l_{f}+l_{ph}+1} \;
    J(l_{f},l_{ph})
    \displaystyle\sum\limits_{m = \pm 1}
    m \,h_{-m} \;
    \displaystyle\sum\limits_{\mu^{\prime} = \pm 1}
    C_{l_{f}l_{ph}l_{ph}}^{m \mu^{\prime}} f_{l_{f}l_{ph}}^{m \mu^{\prime}}(\theta), \\

  \displaystyle\frac{d \,p_{l}^{E}}{\sin{\theta}\,d\theta} & = &
    -i^{l_{f}+l_{ph}} \;
    \sqrt{\displaystyle\frac{l_{ph}+1}{2l_{ph}+1}} \, J(l_{f},l_{ph}-1)
    \displaystyle\sum\limits_{m = \pm 1}
    h_{-m} \;
    \displaystyle\sum\limits_{\mu^{\prime} = \pm 1}
      C_{l_{f},l_{ph},l_{ph}-1}^{m \mu^{\prime}} \: f_{l_{f},l_{ph}-1}^{m \mu^{\prime}}(\theta) \: - \\
  & - &
    i^{l_{f}+l_{ph}} \;
    \sqrt{\displaystyle\frac{l_{ph}}{2l_{ph}+1}} \, J(l_{f},l_{ph}+1)
    \displaystyle\sum\limits_{m = \pm 1}
    h_{-m} \;
    \displaystyle\sum\limits_{\mu^{\prime} = \pm 1}
      C_{l_{f},l_{ph},l_{ph}+1}^{m \mu^{\prime}} \: f_{l_{f},l_{ph}+1}^{m \mu^{\prime}}(\theta),
\end{array}
\label{eq.2.5.4}
\end{equation}
where
\begin{equation}
\begin{array}{lcl}
  C_{l_{f} l_{ph} n}^{m \mu^{\prime}} & = &
    (-1)^{l_{f}+n+1 - \mu^{\prime} + \frac{|m+\mu^{\prime}|}{2}} \;
    (n, 1, l_{ph} \big| -m-\mu^{\prime}, \mu^{\prime}, -m) \; \times \\
  & \times &
    \sqrt{\displaystyle\frac{(2l_{f}+1)\,(2n+1)}{32\pi}\;
          \displaystyle\frac{(l_{f}-1)!}{(l_{f}+1)!} \;
          \displaystyle\frac{(n-|m+\mu^{\prime}|)!}{(n+|m+\mu^{\prime}|)!}}, \\
  f_{l_{f} n}^{m \mu^{\prime}}(\theta) & = &
    P_{l_{f}}^{1}(\cos{\theta}) \; P_{1}^{1}(\cos{\theta}) \; P_{n}^{|m+\mu^{\prime}|} (\cos{\theta}).
\end{array}
\label{eq.2.5.5}
\end{equation}
One can see that integration of functions (\ref{eq.2.5.4}) by angle $\theta$ with limits from 0 to $\pi$ gives the total matrix elements $p_{l}^{M}$ and $p_{l}^{E}$ exactly for transition into superposition of all possible final states with different $m_{f}$ at the same $l_{f}$.



We shall find the matrix element at the first values of $l_{f}$ and $l_{ph}$. We have $l_{f} = 1$, $l_{ph} = 1$. Calculating coefficients $C_{11 n}^{m \mu^{\prime}}$ and functions $f_{1 n}^{m \mu^{\prime}}(\theta)$ (see Appendix~\ref{app.4} and \ref{app.5}), from eq.~(\ref{eq.2.5.4})
we obtain:
\begin{equation}
\begin{array}{lcl}
  \displaystyle\frac{d \,\tilde{p}_{1}^{M}}{\sin{\theta}\,d\theta} & = &
    - \displaystyle\frac{3}{8} \: \sqrt{\displaystyle\frac{1}{\pi}} \cdot
    J(1,1) \cdot
    \sin^{2}{\theta} \cos{\theta}, \\
  \displaystyle\frac{d \,\tilde{p}_{1}^{E}}{\sin{\theta}\,d\theta} & = &
    i \: \displaystyle\frac{1}{8} \: \sqrt{\displaystyle\frac{2}{\pi}} \cdot  J(1,0) \cdot  \sin^{2}{\theta} \: + \:
    i \: \displaystyle\frac{1}{8} \: \sqrt{\displaystyle\frac{1}{\pi}} \cdot  J(1,2) \cdot
    \sin^{2}{\theta} \: \Bigl( 1 - 3 \sin^{2}{\theta} \Bigr).
\end{array}
\label{eq.2.6.1}
\end{equation}
Integrating these expressions over angle $\theta$, we find the integral matrix elements:
\begin{equation}
\begin{array}{lcllcl}
  \tilde{p}_{1}^{M} & = & 0, &
  \tilde{p}_{1}^{E} & = &
    i \: \displaystyle\frac{1}{6} \, \sqrt{\displaystyle\frac{2}{\pi}} \cdot
    \Bigl\{ J(1,0) - \displaystyle\frac{7}{10} \, \sqrt{2} \cdot J(1,2) \Bigr\}.
\end{array}
\label{eq.2.6.2}
\end{equation}

\subsection{Dipole approximation and Fermi's Golden rule
\label{sec.2.7}}

\subsubsection{Vector potential $\mathbf{A}$ and integrated matrix element
\label{sec.2.7.1}}

The matrix element $p\,(k_{i},k_{f})$ in the dipole approximation of the vector potential $\mathbf{A}$ has a form:
\begin{equation}
  p\,(k_{i}, k_{f}) =
    \sum\limits_{\mu = -1, 1}  h_{\mu}\,\mathbf{\xi}^{*}_{\mu}
    \int
      \psi^{*}_{f}(\mathbf{r})\:
      \displaystyle\frac{\partial}{\partial \mathbf{r}}\: \psi_{i}(\mathbf{r})\; \mathbf{dr}.
\label{eq.2.7.1.1}
\end{equation}
Applying the following transformation at $l_{i}=0$ introduced in~\cite{Papenbrock.1998.PRLTA}
(see new analog in a general case in Appendix~\ref{app.7}):
\begin{equation}
\begin{array}{lcl}
  \Bigl\langle f \Bigl|\, \displaystyle\frac{\partial}{\partial \mathbf{r}}\, \Bigl|\, i \Bigl\rangle =
  \displaystyle\frac{1}{w_{fi}}\;
    \biggl\langle f\,
    \biggl|\,
      \displaystyle\frac{\partial\, V(\mathbf{r})}{\partial \mathbf{r}}\,
    \biggr|\, i \biggr\rangle,
\end{array}
\label{eq.2.7.1.2}
\end{equation}
we obtain:
\begin{equation}
  p\,(k_{i}, k_{f}) =
  \displaystyle\frac{1}{w_{fi}}
    \sum\limits_{\mu = -1, 1}  h_{\mu}\,\mathbf{\xi}^{*}_{\mu}
    \int
      \psi^{*}_{f}(\mathbf{r})\:\psi_{i}(\mathbf{r})\;
      \displaystyle\frac{\partial\, V(\mathbf{r})}{\partial \mathbf{r}}\; \mathbf{dr}.
\label{eq.2.7.1.3}
\end{equation}
In the spherically symmetric approximation of the $\alpha$-decay we use the wave functions in form (\ref{eq.2.4.2})
and then the matrix element transforms into
\begin{equation}
  p\,(k_{i}, k_{f}) =
  \displaystyle\frac{1}{w_{fi}}
    \sum\limits_{\mu = -1, 1}
    \sum\limits_{m_{f}}\:
      h_{\mu}\,\mathbf{\xi}^{*}_{\mu}
    \displaystyle\int\limits^{+\infty}_{0} r^{2} dr
    \displaystyle\int d\Omega \cdot
      \varphi_{f}^{*} (r) \: Y_{l_{f}m_{f}}^{*} ({\mathbf n}_{r}^{f}) \cdot
      \varphi_{i} (r) \cdot
      \displaystyle\frac{\partial\, V(r)}{\partial \mathbf{r}}.
\label{eq.2.7.1.4}
\end{equation}
After use of the gradient formula (\ref{eq.2.4.7}) at $l_{i}=0$ and taking into account form (\ref{eq.2.4.11}) for the spherical function $\mathbf{T}_{01,0}(\mathbf{n}_{r})$,
this matrix element is separated into radial and angular integrals:
\begin{equation}
\begin{array}{ccl}
  p\,(k_{i}, k_{f}) & = &
  -\,\displaystyle\frac{1}{w_{fi}}\,
    \sqrt{\displaystyle\frac{1}{3}}
    \displaystyle\sum\limits_{\mu = -1, 1} h_{\mu}
    \displaystyle\sum\limits_{\mu^{\prime} = \pm 1}
      \mathbf{\xi}^{*}_{\mu}\,\mathbf{\xi}_{\mu^{\prime}}
      \displaystyle\int\limits^{+\infty}_{0}
        \varphi_{f}^{*}(r)\;
        \varphi_{i} (r)\: \displaystyle\frac{d\, V(r)}{dr}\;
        r^{2} dr\; \cdot
    \displaystyle\sum\limits_{m_{f}}\:
      \displaystyle\int
        Y_{l_{f}m_{f}}^{*} ({\mathbf n}_{r}^{f})\;
        Y_{1,-\mu^{\prime}}(\mathbf{n}_{r})\; d\Omega
\end{array}
\label{eq.2.7.1.5}
\end{equation}
or ($\mathbf{\xi}_{\pm 1}$ and $\mathbf{\xi}_{\pm 1}^{*}$ are orthogonal vectors)
\begin{equation}
\begin{array}{lcl}
  p\,(k_{i}, k_{f}) =
  -\,\displaystyle\frac{1}{w_{fi}}\,
    J_{\rm dip}\, (l_{f}) \cdot
    I_{\rm dip}\, (l_{f}),
\end{array}
\label{eq.2.7.1.6}
\end{equation}
where
\begin{equation}
\begin{array}{lcl}
  J_{\rm dip}\, (l_{f}) & = &
    \displaystyle\int\limits^{+\infty}_{0}
      \varphi_{f}^{*}(l_{f}, r)\;
      \varphi_{i} (r)\;
      \displaystyle\frac{d\, V(r)}{dr}\;
      r^{2} dr, \\

  I_{\rm dip}\, (l_{f}) & = &
    \sqrt{\displaystyle\frac{1}{3}}\;
    \displaystyle\sum\limits_{\mu = \pm 1}
    \displaystyle\sum\limits_{m}
      h_{\mu}
    \displaystyle\int
        Y_{l_{f}m_{f}}^{*} ({\mathbf n}_{r}^{f})\;
        Y_{1,-\mu}(\mathbf{n}_{r})\; d\Omega.
\end{array}
\label{eq.2.7.1.7}
\end{equation}
After use of formula (\ref{eq.2.5.1}) for vectors $\mathbf{n}^{f}_{r}$ and $\mathbf{n}_{r}$
the angular integral $I_{\rm dip}\, (l_{f})$ is non-zero only at
\begin{equation}
\begin{array}{lcl}
  m_{f} = -\mu = \pm 1, &
  l_{f} = 1
\end{array}
\label{eq.2.7.1.8}
\end{equation}
and equals to
\begin{equation}
\begin{array}{lcl}
  I_{\rm dip}\, (l_{f}=1) & = &
  \sqrt{\displaystyle\frac{1}{3}}\; \displaystyle\sum\limits_{\mu = \pm 1} h_{\mu} =
  \sqrt{\displaystyle\frac{1}{3}}\; \Bigl(h_{-1} + h_{+1}\Bigr) =
  \sqrt{\displaystyle\frac{1}{3}}\; \Bigl(-i\,\sqrt{2}\Bigr) =
  -i\; \sqrt{\displaystyle\frac{2}{3}}\; .
\end{array}
\label{eq.2.7.1.9}
\end{equation}
Now the matrix element (\ref{eq.2.7.1.6}) obtains a form:
\begin{equation}
\begin{array}{lcl}
  p\,(k_{i}, k_{f}) =
  i\; \sqrt{\displaystyle\frac{2}{3}}\; \displaystyle\frac{J_{\rm dip}\, (1)}{w_{fi}}.
\end{array}
\label{eq.2.7.1.10}
\end{equation}
Comparing selection rules~(\ref{eq.2.7.1.8}) in the dipole approximation with the selection rules (\ref{eq.2.5.3}) in the multipole approach, one can see that they do not impose any restrictions on the emission of photons.
\emph{From eq.~(\ref{eq.2.7.1.1}) one can see that the wave function of photon in the dipole approximation has no any information about orientation of the vector $\mathbf{k}$ concerning vector $\mathbf{r}$ in any selected frame system. On the basis of such a fact it is logically to consider the emission of photons in the dipole approximation as isotropic.}



\subsubsection{Differential matrix element
\label{sec.2.7.2}}

Let us define the following differential components of the angular integral $I_{\rm dip}$
by solid angle $\Omega$ and by angle $\theta$:
\begin{equation}
\begin{array}{lcl}
  \displaystyle\frac{d\, I_{\rm dip}\, (l_{f})}{d\Omega} & = &
  \displaystyle\frac{d\, I_{\rm dip}\, (l_{f})}{\sin{\theta}\,d\theta\, d\varphi} =
    \sqrt{\displaystyle\frac{1}{3}}\;
    \displaystyle\sum\limits_{\mu = \pm 1}
    \displaystyle\sum\limits_{m}
      h_{\mu} \cdot
        Y_{l_{f}m_{f}}^{*} ({\mathbf n}_{r})\;
        Y_{1,-\mu}(\mathbf{n}_{r}), \\

  \displaystyle\frac{d\, I_{\rm dip}\, (l_{f})}{\sin{\theta}\,d\theta} & = &
    \sqrt{\displaystyle\frac{1}{3}}\;
    \displaystyle\sum\limits_{\mu = \pm 1}
    \displaystyle\sum\limits_{m}
      h_{\mu} \cdot
    \displaystyle\int\limits_{0}^{2\pi}
        Y_{l_{f}m_{f}}^{*} ({\mathbf n}_{r})\;
        Y_{1,-\mu}(\mathbf{n}_{r})\; d\varphi.
\end{array}
\label{eq.2.7.2.1}
\end{equation}
But, in their calculations we shall assume that the selection rules (\ref{eq.2.7.1.8}) are fulfilled.
We find:
\begin{equation}
\begin{array}{lcl}
  \displaystyle\frac{d\, I_{\rm dip}\, (l_{f})}{d\Omega} & = &
    \sqrt{\displaystyle\frac{1}{3}}\;
      \displaystyle\sum\limits_{\mu = \pm 1} h_{\mu} \cdot
      \displaystyle\frac{3}{8\pi} \cdot
      P_{1}^{1} (\cos{\theta}) \cdot P_{1}^{1} (\cos{\theta}) =
    -i\; \displaystyle\frac{\sqrt{6}}{8\pi}\; \sin^{2}{\theta}, \\

  \displaystyle\frac{d\, I_{\rm dip}\, (l_{f})}{\sin{\theta}\,d\theta} & = &
  \displaystyle\int\limits_{0}^{2\pi}
    \displaystyle\frac{d\, I_{\rm dip}\, (l_{f})}{d\Omega}\; d\varphi =
    -i\; \displaystyle\frac{\sqrt{6}}{4}\; \sin^{2}{\theta}.
\end{array}
\label{eq.2.7.2.2}
\end{equation}
On their basis we shall define the following differential components of the matrix element:
\begin{equation}
\begin{array}{lcl}
  \displaystyle\frac{d\, p_{\rm dip}\, (l_{f}=1)}{d\Omega} & = &
  -\,\displaystyle\frac{J_{\rm dip}\,(1)}{w_{fi}} \cdot \displaystyle\frac{d\, I_{\rm dip}\, (l_{f})}{d\Omega} =
  i\;\displaystyle\frac{\sqrt{6}}{8\pi}\;
    \displaystyle\frac{J_{\rm dip}\, (1)}{w_{fi}} \cdot \sin^{2}{\theta}, \\
  \displaystyle\frac{d\, p_{\rm dip}\, (l_{f}=1)}{\sin{\theta}\,d\theta} & = &
  -\,\displaystyle\frac{J_{\rm dip}\, (1)}{w_{fi}} \cdot
    \displaystyle\frac{d\, I_{\rm dip}\, (l_{f})}{\sin{\theta}\,d\theta} =
  i\;\displaystyle\frac{\sqrt{6}}{4}\; \displaystyle\frac{J_{\rm dip}\, (1)}{w_{fi}} \cdot \sin^{2}{\theta}.
\end{array}
\label{eq.2.7.2.3}
\end{equation}
One can see that integration of such functions over the solid angle $\Omega$ or over the angle $\theta$ (with needed limits) gives the integrated matrix element $p_{\rm dip}$ exactly. Comparing eqs.~(\ref{eq.2.7.2.3}) and (\ref{eq.2.7.1.10}), we obtain connection between the differential and integral components of the matrix element
in the dipole approximation:
\begin{equation}
\begin{array}{lclcl}
  \displaystyle\frac{d\, p_{\rm dip}}{d\Omega} & = &
  2\pi \cdot \displaystyle\frac{d\, p_{\rm dip}}{\sin{\theta}\,d\theta} & = &
  p_{\rm dip} \cdot
    \displaystyle\frac{3}{8\pi}\;
    \sin^{2}{\theta}.
\end{array}
\label{eq.2.7.2.4}
\end{equation}
The differential matrix element characterizes the angular distribution of emitted $\alpha$-particles, which becomes anisotropic in result of emission of photons.



\subsection{Angular probability of emission of photon with impulse $\mathbf{k}$ and polarization $\mathbf{e}^{(\alpha)}$
\label{sec.2.8}}

I define the probability of transition of the system for time unit from the initial $i$-state into the final $f$-states, being in the given interval $d \nu_{f}$, with emission of photon with possible impulses inside the given interval $d \nu_{ph}$, so (see ref.\cite{Landau.v3.1989}, (42,5) \S~42, p.~189; ref.\cite{Berestetsky.1989}, \S~44, p.~191):
\begin{equation}
\begin{array}{lll}
  d W = \displaystyle\frac{|a_{fi}|^{2}}{T} \cdot d\nu =
    2\pi \:|F_{fi}|^{2} \: \delta (w_{f} - w_{i} + w) \cdot d\nu, &
  d \nu = d\nu_{f} \cdot d\nu_{ph}, &
  d \nu_{ph} = \displaystyle\frac{d^{3} k}{(2\pi)^{3}} =
          \displaystyle\frac{w^{2} \, dw \,d\Omega_{ph}}{(2\pi c)^{3}},
\end{array}
\label{eq.2.8.1}
\end{equation}
where $d\nu_{ph}$ and $d\nu_{f}$ are intervals defined for photon and particle in the final $f$-state,
$d\Omega_{ph} = d\,\cos{\theta_{ph}} = \sin{\theta_{ph}} \,d\theta_{ph} \,d\varphi_{ph}$, $k_{ph}=w/c$.
$F_{fi}$ is integral over space with possible summation by some quantum numbers of the system in the final $f$-state. Such procedure is averaging by these characteristics and $F_{fi}$ is independent on them. Then, interval $d\,\nu_{f}$ has only new characteristics and quantum numbers, by which integration and summation in $F_{fi}$ was not fulfilled.
Integrating eq.~(\ref{eq.2.8.1}) over $dw$ and substituting eq.~(\ref{eq.2.1.2}) for $F_{fi}$, we find:
\begin{equation}
\begin{array}{cc}
  d W = \displaystyle\frac{Z_{eff}^{2} \,e^{2}}{m^{2}}\:
        \displaystyle\frac{\hbar\, w_{fi}}{2\pi \,c^{3}} \; \Bigl|p(k_{i}, k_{f})\Bigr|^{2} \;
        d \Omega_{ph} \, d\nu_{f}, &
  w_{fi} = w_{i} - w_{f} = \displaystyle\frac{E_{i} - E_{f}}{\hbar}.
\end{array}
\label{eq.2.8.2}
\end{equation}
This is the probability of the photon emission with impulse $\mathbf{k}$ (and with averaging by polarization $\mathbf{e}^{(\alpha)}$) where the integration over angles of the particle motion after the photon emission has already fulfilled.

To take direction $\mathbf{n}_{\mathbf{r}}^{f}$ of motion (or tunneling) of the particle after emission into account, I define the probability so:
\emph{the \underline{angular probability} concerning angle $\theta$ is such a function, definite integral of which by the angle $\theta$ with limits from 0 to $\pi$ corresponds exactly to the total probability of photon emission (\ref{eq.2.8.2})}:
\begin{equation}
\begin{array}{ccl}
  \displaystyle\frac{d W(\theta_{f})} {d\,\Omega_{ph} \: d\cos{\theta_{f}}} & = &
  \displaystyle\frac{Z_{eff}^{2}\, \hbar\, e^{2}}{2\pi\, c^{3}}\: \displaystyle\frac{w_{fi}}{m^{2}} \;
    \biggl\{p\,(k_{i},k_{f}) \displaystyle\frac{d\, p^{*}(k_{i},k_{f}, \theta_{f})}{d\cos{\theta_{f}}} + {\rm h. e.} \biggr\}.
\end{array}
\label{eq.2.8.3}
\end{equation}
This probability is inversely proportional to normalized volume $V$. With a purpose to have the probability independent on $V$, I divide eq.~(\ref{eq.2.8.3}) on flux $j$ of outgoing $\alpha$-particles, which is inversely proportional to this volume $V$ also. Using quantum field theory approach
(where $v(\mathbf{p}) = |\mathbf{p}| / p_{0}$ at $c=1$, see~\cite{Bogoliubov.1980}, \S~21.4, p.~174):
\begin{equation}
\begin{array}{cc}
  j = n_{i}\, v(\mathbf{p}_{i}), &
  v_{i} = |\mathbf{v}_{i}| = \displaystyle\frac{c^{2}\,|\mathbf{p}_{i}|} {E_{i}} =
          \displaystyle\frac{\hbar\,c^{2}\,k_{i}} {E_{i}},
\end{array}
\label{eq.2.8.4}
\end{equation}
where $n_{i}$ is average number of particles in time unit before photon emission (we have $n_{i}=1$ for the normalized wave function in the initial $i$-state), $v(\mathbf{p}_{i})$ is module of velocity of outgoing particle in the frame system where colliding center is not moved, I obtain the \emph{differential absolute probability}
(while let's name $dW$ as the \emph{relative probability}):
\begin{equation}
\begin{array}{ccl}
  \displaystyle\frac{d\,P (\varphi_{f}, \theta_{f})}{d\Omega_{ph}\, d\cos{\theta_{f}}} & = &
  \displaystyle\frac{d\,W (\varphi_{f}, \theta_{f})}{d\Omega_{ph}\, d\cos{\theta_{f}}} \cdot
    \displaystyle\frac{E_{i}} {\hbar\, c^{2}\, k_{i}} =
    \displaystyle\frac{Z_{eff}^{2} \,e^{2}}{2\pi\,c^{5}}\:
      \displaystyle\frac{w_{ph}\,E_{i}}{m^{2}\,k_{i}} \;
      \biggl\{p\,(k_{i},k_{f}) \displaystyle\frac{d\, p^{*}(k_{i},k_{f}, \Omega_{f})}{d\,\cos{\theta_{f}}} + {\rm h. e.} \biggr\}.
\end{array}
\label{eq.2.8.5}
\end{equation}

\subsection{Multipolar approach
\label{sec.2.9}}

Let us find the bremsstrahlung probability in the multiple approach at the first values $l_{f}=1$ and $l_{ph}=1$. Starting from eqs.~(\ref{eq.2.4.3}) and (\ref{eq.2.4.4}), and using the found differential and integral electrical and magnetic components (\ref{eq.2.6.1}) and (\ref{eq.2.6.2}),
I calculate:
\begin{equation}
\begin{array}{lcl}
\vspace{2mm}
  \tilde{p}_{1}\, (k_{i},k_{f}) & = &
    - i \, \sqrt{\displaystyle\frac{1}{3}} \cdot
    \Bigl\{ J(1,0) - \displaystyle\frac{7}{10} \, \sqrt{2} \cdot J(1,2) \Bigr\}, \\

  \displaystyle\frac{d \, \tilde{p}_{1}\, (k_{i},k_{f})}{\sin{\theta}\,d\theta} & = &
    i\; \displaystyle\frac{\sqrt{6}}{8} \: \cdot
    \biggl\{
      3\,J(1,1) \cdot \cos{\theta} - \sqrt{2}\, J(1,0) - J(1,2) \cdot \Bigl( 1 - 3 \sin^{2}{\theta} \Bigr)
    \biggr\} \cdot \sin^{2}{\theta}
\end{array}
\label{eq.2.9.1}
\end{equation}
and from eq.~(\ref{eq.2.8.5}) I obtain the \emph{absolute} angular probability
($\vartheta_{\alpha\gamma} \equiv \theta_{f}$):
\begin{equation}
\begin{array}{ccl}
\vspace{3mm}
  \displaystyle\frac{d P^{E1+M1}_{1}(\vartheta_{\alpha\gamma})}
    {d\,\Omega_{ph} \: d\cos{\vartheta_{\alpha\gamma}}} & = &
    \displaystyle\frac{Z_{eff}^{2}\, e^{2}}{8\,\pi\, c^{5}}\:
    \displaystyle\frac{w_{fi}}{m^{2}}\,
    \displaystyle\frac{E_{i}}{k_{i}}\;
    \biggl\{ \Bigl[ J(1,0) - \displaystyle\frac{7}{10} \, \sqrt{2} \cdot J(1,2) \Bigr] \times \\
  & \times &
      \Bigl[
        J^{*}(1,0) +
        \displaystyle\frac{1}{\sqrt{2}} J^{*}(1,2) \cdot \Bigl( 1 - 3 \sin^{2}{\vartheta_{\alpha\gamma}} \Bigr)
        - \displaystyle\frac{3}{\sqrt{2}} \,J^{*}(1,1) \cdot \cos{\vartheta_{\alpha\gamma}}
      \Bigr]
    + {\rm h. e.} \biggr\} \cdot \sin^{2}{\vartheta_{\alpha\gamma}}.
\end{array}
\label{eq.2.9.2}
\end{equation}

\subsection{Dipole approach
\label{sec.2.10}}

Let us find the differential absolute probability (\ref{eq.2.8.5}) in dependence on angle $\theta$ and the integrated absolute probability in the dipole approximation.
Taking into account:
\[
\begin{array}{cclccl}
  p_{\rm dip}\, (l_{f}=1)& = &
    i\;\sqrt{\displaystyle\frac{2}{3}}\; \displaystyle\frac{J_{\rm dip}\, (1)}{w_{fi}}, &
  \hspace{10mm}
  \displaystyle\frac{d\, p_{\rm dip}\, (l_{f}=1)}{\sin{\theta}\,d\theta} & = &
    i\;\displaystyle\frac{\sqrt{6}}{4}\; \displaystyle\frac{J_{\rm dip}\, (1)}{w_{fi}} \cdot \sin^{2}{\theta},
\end{array}
\]
I obtain:
\begin{equation}
\begin{array}{ccl}
\vspace{3mm}
  \displaystyle\frac{d P_{\rm dip}(\theta_{f})} {d\,\Omega_{ph} \: d\cos{\theta_{f}}} & = &
    \displaystyle\frac{Z_{\rm eff}^{2}\, e^{2}}{c^{5}}\:
    \displaystyle\frac{E_{i}}{m^{2}\,k_{i}\,w_{fi}} \cdot
    \Bigl| J_{\rm dip}\, (1) \Bigr|^{2} \cdot \sin^{2}{\theta}, \\

  P_{\rm dip}(\theta_{f}) & = &
    \displaystyle\frac{Z_{\rm eff}^{2}\, e^{2}}{3\pi\,c^{5}}\: \displaystyle\frac{E_{i}}{m^{2}\,k_{i}\,w_{fi}} \cdot
    \Bigl| J_{\rm dip}\, (1) \Bigr|^{2}.
\end{array}
\label{eq.2.10.1}
\end{equation}
One can write:
\begin{equation}
\begin{array}{ccl}
\vspace{3mm}
  \displaystyle\frac{d P_{\rm dip}(\theta_{f})} {d\,\Omega_{ph} \: d\cos{\theta_{f}}} & = &
  3\pi \cdot P_{\rm dip}(\theta_{f}) \cdot \sin^{2}{\theta}.
\end{array}
\label{eq.2.10.2}
\end{equation}
Comparing angular dependence in such a result with the angular correlation obtained in the dipole approximation in~\cite{Jentschura.2008.PRC} (see eq.~(9) in the cited paper) one can find that they coincide.


\subsection{Spectroscopic factor
\label{sec.2.11}}

In order to take into account a non-unit possibility of formation of the $\alpha$-particle in the state, from which it is further emitted outside, let us come to other problems of nuclear decays where this question has already been resolved. For example, in a problem of decay of nucleus through emission of proton we should take into account that the state which is occupied by the proton before its emission, is empty for the daughter nucleus.
In order to obtain proper value for half-life we should divide it, calculated before by semiclassical approach (or others) directly, on the spectroscopic factor $S_{p}^{\rm (th)}$ (see \cite{Aberg.1997.PRC}).
The spectroscopic factor can easily be calculated in the independent quasiparticle approximation (BCS), in which one assume that the ground state of odd-Z nucleus is one-quasiparticle state, while that of odd-odd system is two-quasiparticle configuration. In the BSC theory, the spectroscopic factor is given by $S^{\rm (th)}_{p} = u^{2}$, where $u^{2}$ is the probability that the spherical orbital corresponding to proton emitted is empty in the daughter nucleus \cite{Aberg.1997.PRC,Heyde.1990}. For different proton emitters inclusion of the spectroscopic factors into formulas of half-lives improves agreement between calculated half-lives and their experimental values. Similar approach seems to be in the problem of $\alpha$-decay, where the spectroscopic factor $S_{\alpha}^{\rm (th)}$ is successfully applied.

On such a basis, let us modify the formulas for probability. For the multipole and dipole approaches from eqs.~(\ref{eq.2.9.2}) and~(\ref{eq.2.10.1}) we obtain
the following \emph{modified absolute angular probabilities}:
\begin{equation}
\begin{array}{ccl}
\vspace{3mm}
  \displaystyle\frac{d \tilde{P}^{E1+M1}_{{\rm mult},\,1}(\vartheta_{\alpha\gamma})}
    {d\,\Omega_{ph} \: d\cos{\vartheta_{\alpha\gamma}}} & = &
    \displaystyle\frac{Z_{eff}^{2}\, e^{2}}{8\,\pi\, c^{5}\: S_{\alpha}^{\rm (th)}}\:
    \displaystyle\frac{w_{fi}}{m^{2}}\,
    \displaystyle\frac{E_{i}}{k_{i}}\;
    \biggl\{ \Bigl[ J(1,0) - \displaystyle\frac{7}{10} \, \sqrt{2} \cdot J(1,2) \Bigr] \times \\
  & \times &
      \Bigl[
        J^{*}(1,0) +
        \displaystyle\frac{1}{\sqrt{2}} J^{*}(1,2) \cdot \Bigl( 1 - 3 \sin^{2}{\vartheta_{\alpha\gamma}} \Bigr)
        - \displaystyle\frac{3}{\sqrt{2}} \,J^{*}(1,1) \cdot \cos{\vartheta_{\alpha\gamma}}
      \Bigr]
    + {\rm h. e.} \biggr\} \cdot \sin^{2}{\vartheta_{\alpha\gamma}},
\end{array}
\label{eq.2.11.1}
\end{equation}
\begin{equation}
\begin{array}{ccl}
\vspace{3mm}
  \displaystyle\frac{d \tilde{P}_{\rm dip}(\theta_{f})} {d\,\Omega_{ph} \: d\cos{\theta_{f}}} & = &
    \displaystyle\frac{Z_{\rm eff}^{2}\, e^{2}}{c^{5}\: S_{\alpha}^{\rm (th)}}\:
    \displaystyle\frac{E_{i}}{m^{2}\,k_{i}\,w_{fi}} \cdot
    \Bigl| J_{\rm dip}\, (1) \Bigr|^{2} \cdot \sin^{2}{\theta}, \\
  \tilde{P}_{\rm dip}(\theta_{f}) & = &
    \displaystyle\frac{Z_{\rm eff}^{2}\, e^{2}}{3\pi\,c^{5}\: S_{\alpha}^{\rm (th)}}\: \displaystyle\frac{E_{i}}{m^{2}\,k_{i}\,w_{fi}} \cdot
    \Bigl| J_{\rm dip}\, (1) \Bigr|^{2}.
\end{array}
\label{eq.2.11.2}
\end{equation}
One can see that inclusion of the spectroscopic factor $S_{\alpha}^{\rm (th)}$ into formulas of the bremsstrahlung probability raises the spectra obtained before without it by eqs.~(\ref{eq.2.9.2}) and~(\ref{eq.2.10.1}). However, in this paper we shall restrict ourselves only by the first preliminary calculations of the bremsstrahlung spectra for the deformed $^{226}{\rm Ra}$ nucleus.


\section{Calculations and analysis
\label{sec.3}}

\begin{figure}[htbp]
\centerline{%
\includegraphics[width=44mm]{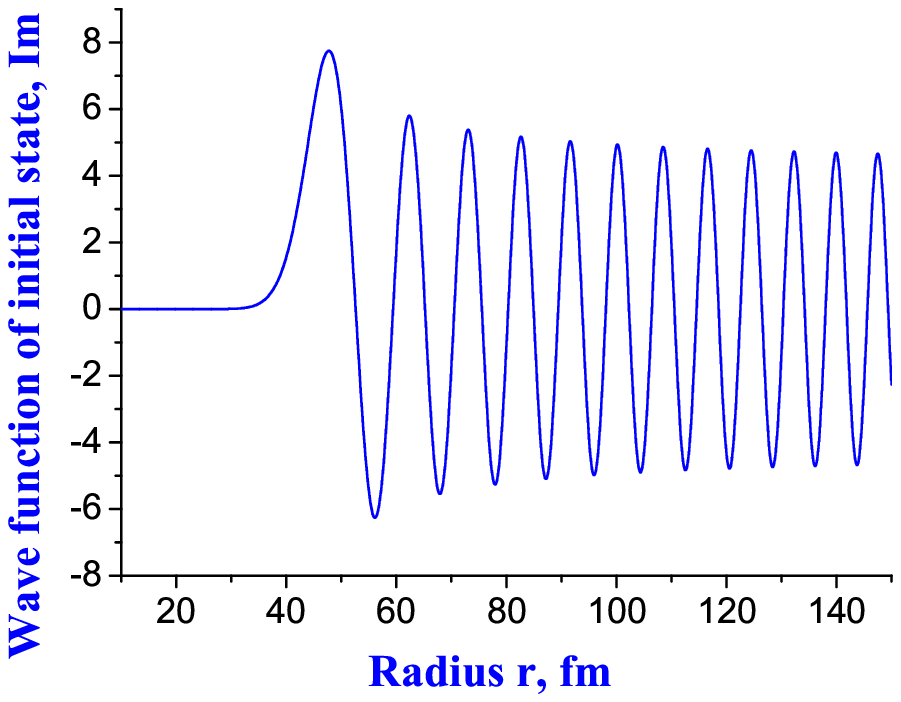}
\includegraphics[width=44mm]{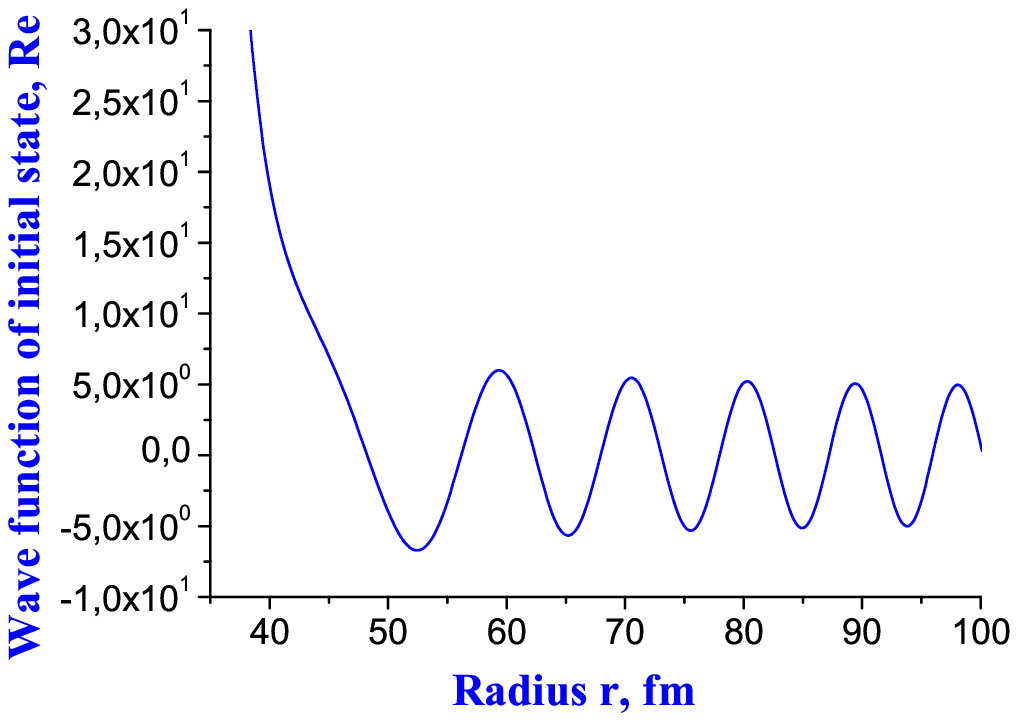}
\includegraphics[width=44mm]{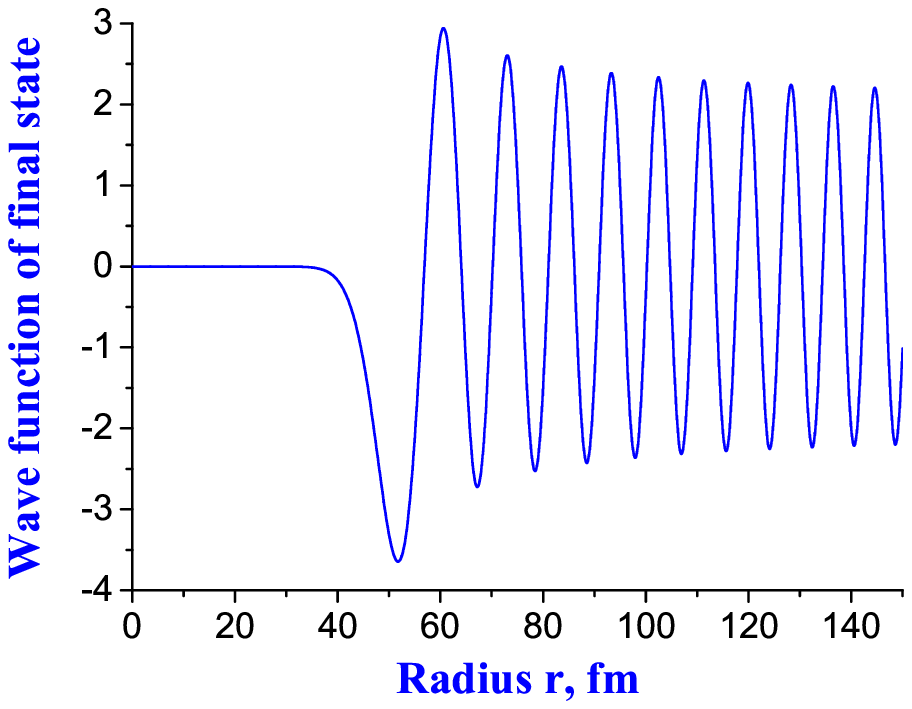}
\includegraphics[width=44mm]{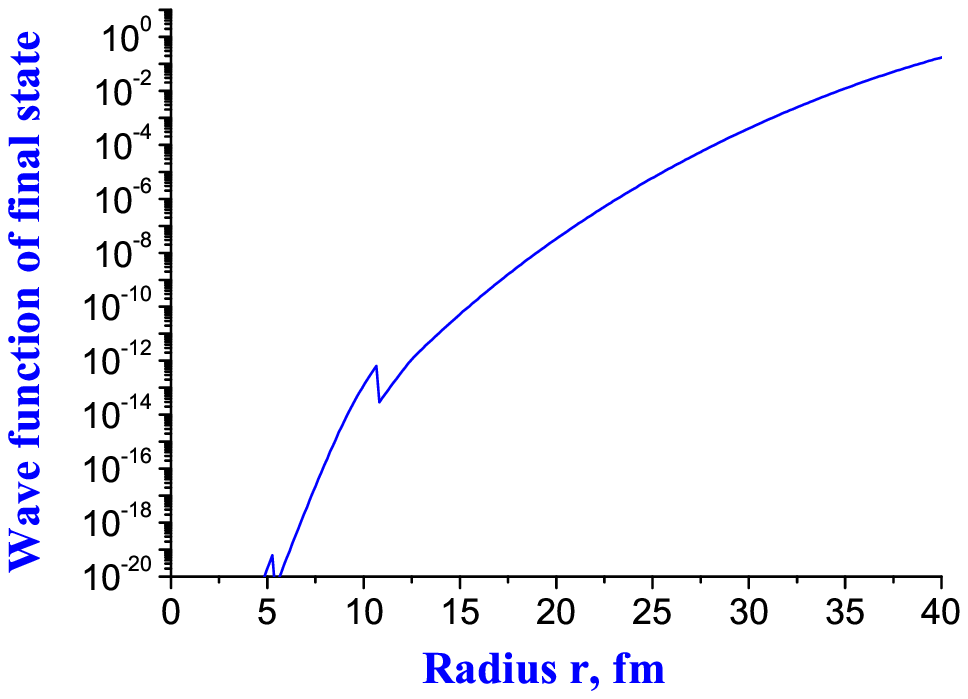}}
\vspace{-4mm}
\caption{Wave functions of the $\alpha$-decaying system for the $^{210}{\rm Po}$ nucleus:
(a) Imaginary part of the wave function in the initial $i$-state;
(b) Real part of the wave function in the initial $i$-state;
(c) The real wave function in the final $f$-state (after emission of photon);
(d) Errors of the wave function in the final $f$-state appeared in current calculations of the spectra (for presentation, module of such a wave function is shown).
\label{fig.1}}
\end{figure}

In order to estimate efficiency of the model and accuracy, which it gives in determination of the angular absolute probability of the photon emission, I have calculated the spectra for the $^{210}\mbox{\rm Po}$, $^{214}\mbox{\rm Po}$, $^{226}\mbox{\rm Ra}$ and $^{244}{\rm Cm}$ nuclei. Here, the bremsstrahlung probability is calculated by eq.~(\ref{eq.2.9.2}).
The nucleus--$\alpha$-particle potential is defined by eqs.~(11)--(15) with parameters --- by eqs.~(16)--(22) in \cite{Maydanyuk.2008.EPJA}.
$Q_{\alpha}$-value is 5.439~MeV for $^{210}\mbox{\rm Po}$, 7.865~MeV for $^{214}\mbox{\rm Po}$, 4.904~MeV for $^{226}\mbox{\rm Ra}$, 5.940~MeV for $^{244}{\rm Cm}$, according to ref.~\cite{Buck.1993.ADNDT} (see p.~63).
The angle $\vartheta_{\alpha\gamma}$ between the directions of the $\alpha$-particle motion
(with possible tunneling) and the photon emission for $^{214}\mbox{\rm Po}$ and $^{226}\mbox{\rm Ra}$ is $90^{\circ}$, that is explained by configuration in experiments \cite{D'Arrigo.1994.PHLTA,Maydanyuk.2008.EPJA,Maydanyuk.2008.MPLA}. For $^{210}{\rm Po}$ I use the same angle but results could be easily generalized for other its values.
For $^{244}{\rm Cm}$ I study two cases at $90^{\circ}$ and $25^{\circ}$ in order to include experiments~\cite{Kasagi.1997.PRLTA} into the total picture.
The first step is calculations of wave functions which should be obtained with high accuracy in order to obtain the convergent spectra. The wave functions of the $\alpha$-decaying system in the states before and after emission of photon are presented in Fig.~\ref{fig.1}
(presentation is for the $^{210}{\rm Po}$ nucleus, energy of photon is 500~keV).
One can see that in the initial state the wave function is complex. This provides flux to be constant (in the internal region, in the region of tunneling and in the external region) and be directed outside. In current calculations of the spectra the wave function in the final state is real that allows to miss discontinuity at $r=0$. In particular, in these figures (a, b, c) one can see behavior of these wave functions at boundary between tunneling and external regions. The last figure (d) shows errors which calculations give us, and one hope this allows us to obtain convergent pictures of the bremsstrahlung probability for all studied nuclei.

\begin{figure}[htbp]
\centerline{%
\includegraphics[width=95mm]{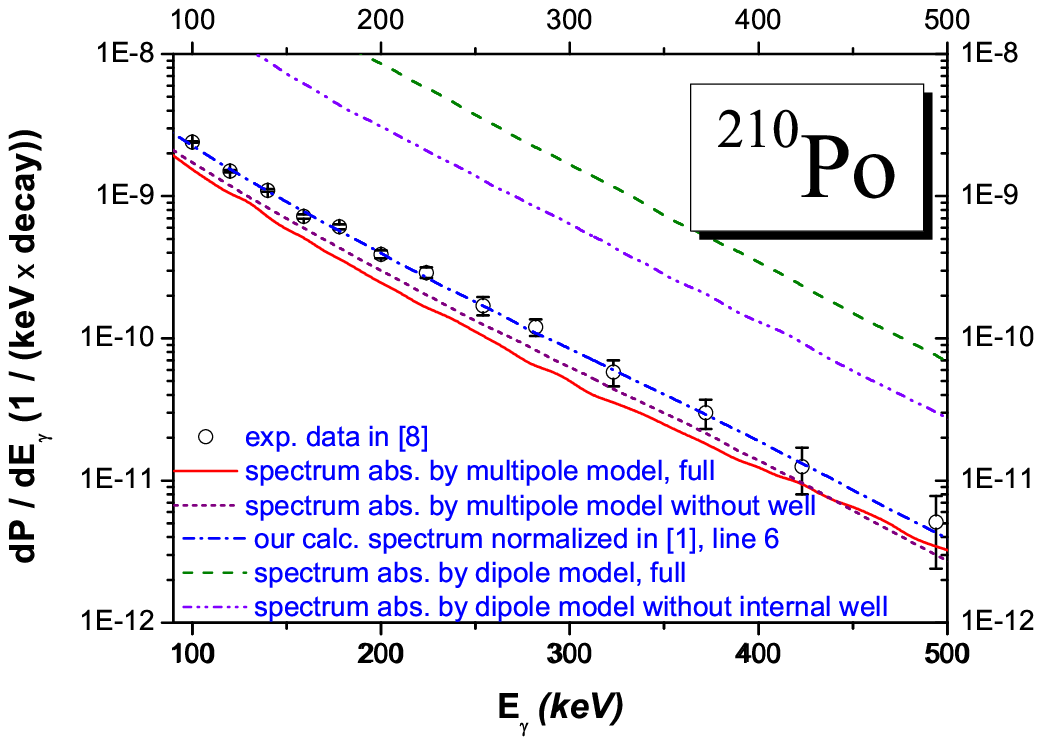}
\hspace{-8mm}\includegraphics[width=95mm]{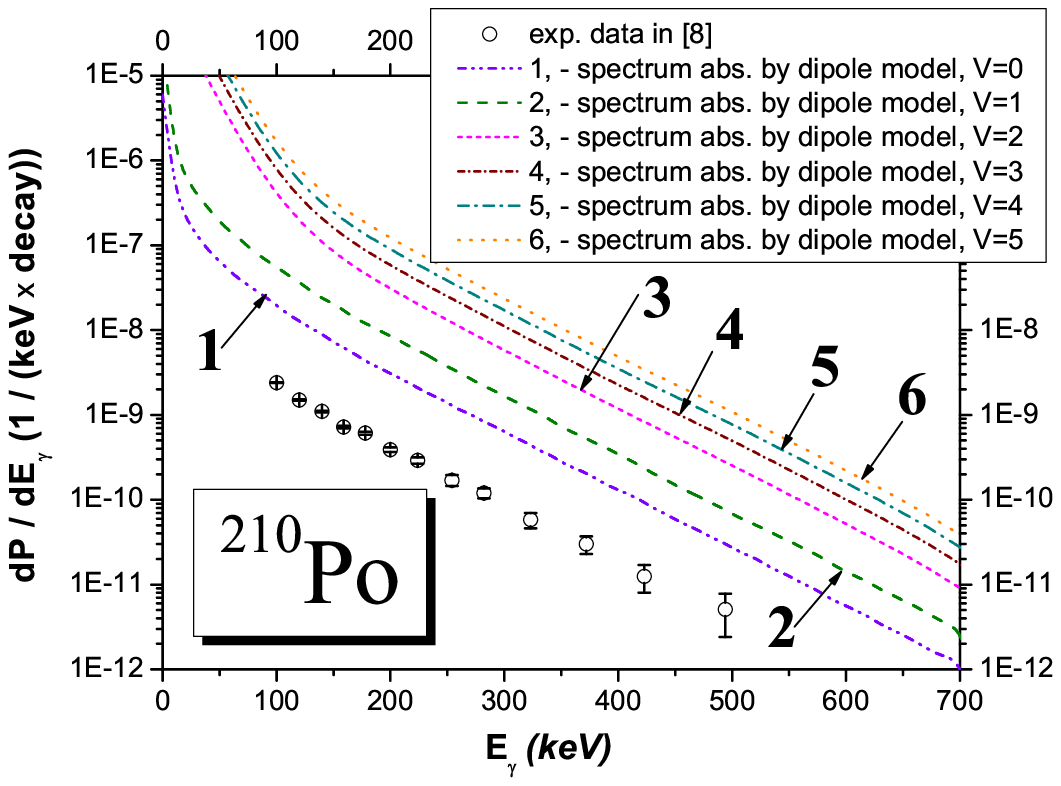}}
\vspace{-4mm}
\caption{Bremsstrahlung probabilities in the $\alpha$-decay of the $^{210}{\rm Po}$ nucleus calculated in the absolute scale
(solid line, red, is for the spectrum calculated by the multipole model;
short-dash line, brown, for the spectrum calculated by the multipole model without emission from the internal region up to the internal turning point;
dash line, green, for the full spectrum calculated by the dipole model;
dash-dot-dot line, violet, for the spectrum calculated by the dipole model without taking the internal region up to the internal turning point into account;
dash-dot line, blue, for the normalized spectrum calculated by the approach in \cite{Maydanyuk.2006.EPJA}
with normalization used in that paper):
[left panel] difference between two spectra calculated in the dipole approach with inclusion of emission from the internal well and without such emission is not small, in contrast to the multipole approach;
[right panel] a role of the nuclear strength $\tilde{V}_{0}$ is clearly shown in calculations of the spectra in the dipole approach: one can see that the strength is larger, the spectrum is higher (for each spectrum the corresponding strength is defined as $\tilde{V}_{0} = V_{0} \cdot V$ where $V_{0}$ is defined by eq.~(14) in~\cite{Denisov.2005.PHRVA} and additional new factor $V$ has values 0,1,2,3,4,5 as shown in the figure).
\label{fig.2}}
\end{figure}

According to analysis, the multipolar approach provides more accurate calculations of the bremsstrahlung spectra both in absolute scale (i.~e. without any normalization on existed experimental data) and in a case when such normalization is used, in comparison with the dipole approach (see Appendix~\ref{app.6}, for details). For example, let us look on Fig.~\ref{fig.2} [left panel], where the spectra for the $^{210}{\rm Po}$ nucleus obtained by the multipole and
dipole approaches in absolute scale are presented.
In order to feel sensitivity of the spectra on change of the internal well of the $\alpha$-nucleus potential (i.~e. its shape in the internal region from $r \to 0$ up to internal turning point), the following calculations are included into this figure:
(1) the spectra without emission from such internal well (for the dipole approach this corresponds to a case of rectangular well in the internal region), and
(2) the complete spectra obtained concerning full $\alpha$-nucleus potential with realistic nuclear component.
It is clearly seen that for the dipole approach the influence of the shape of the potential in the internal region on the spectrum is stronger essentially while the multipole approach seems to be less sensitive to it.
Such a peculiarity could be explained by more accurate use of far asymptotic region by the multipole approach, while the dipole approach reduces such a region.
This leads to higher convergence of numerical integration over $r$ in calculations of the spectra in the dipole approach.
However, \emph{more accurate use of the far asymptotic region by the multipole approach provides essentially better agreement between obtained spectra and experimental data, while the dipole approach seems to do not calculate correctly here (at its present development)}.
This important point adds the power of predictions in absolute scale to the multipolar model, in contrast to the dipole one. So, we shall use the multipolar approach for further estimation of the spectra for interesting nuclei and for analysis.
In order to complete such a picture, I add calculations of the spectra for $^{210}{\rm Po}$ in the dipole approach for different values of the nuclear strength, presented in next Fig.~\ref{fig.2} [right panel]. From here it is clearly seen that the spectrum is higher, if the nuclear strength is larger (i.e. the internal well of the $\alpha$-nucleus potential is deeper). One can suppose that for multipole approach such dependence should be essentially smaller.


\subsection{Bremsstrahlung spectra for $^{214}\mbox{\rm Po}$, $^{226}\mbox{\rm Ra}$ and $^{244}\mbox{\rm Cm}$: comparison theory and experiments
\label{sec.3.3}}

\begin{figure}[htbp]
\centerline{%
\includegraphics[width=62mm]{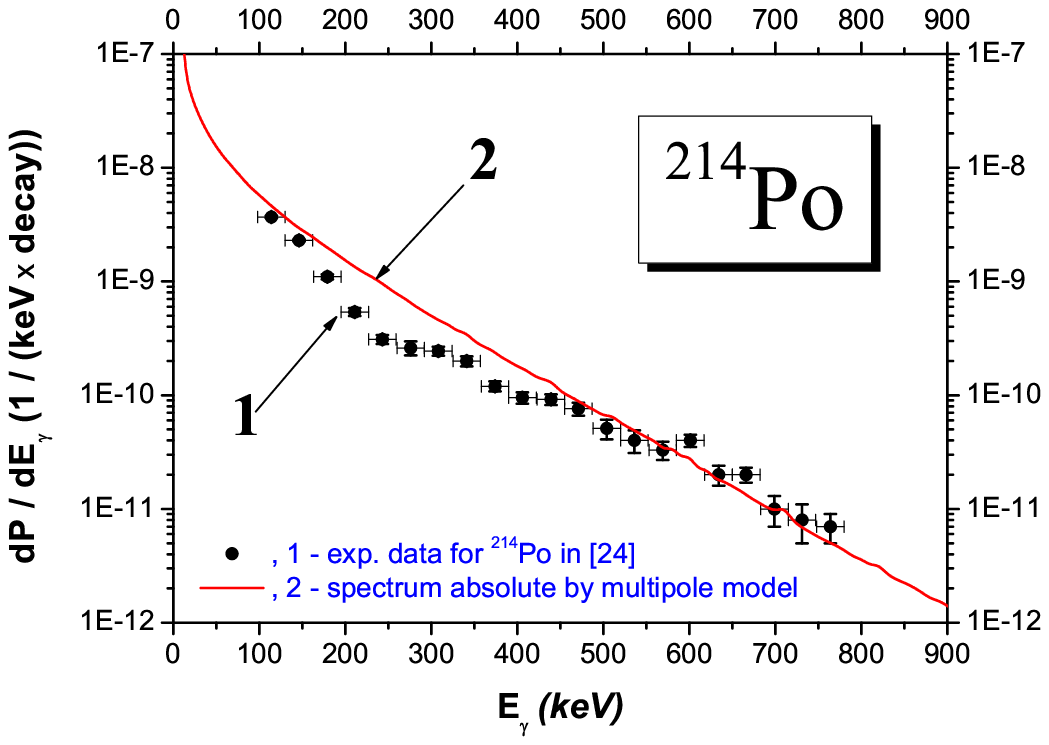}
\hspace{-2mm}\includegraphics[width=62mm]{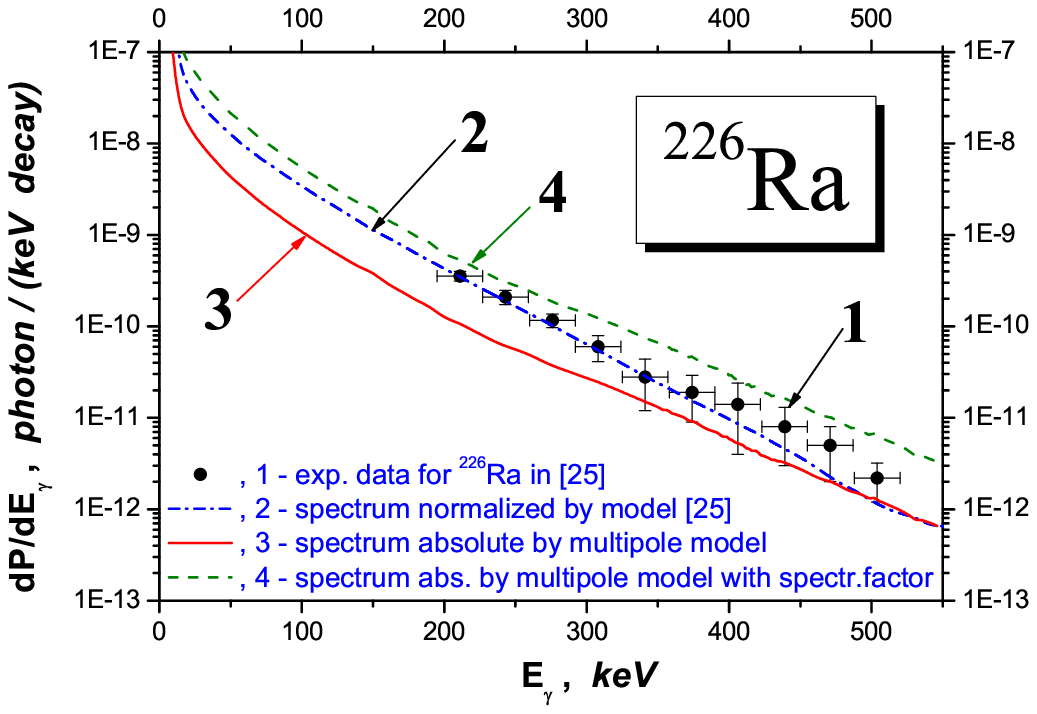}
\hspace{-2mm}\includegraphics[width=62mm]{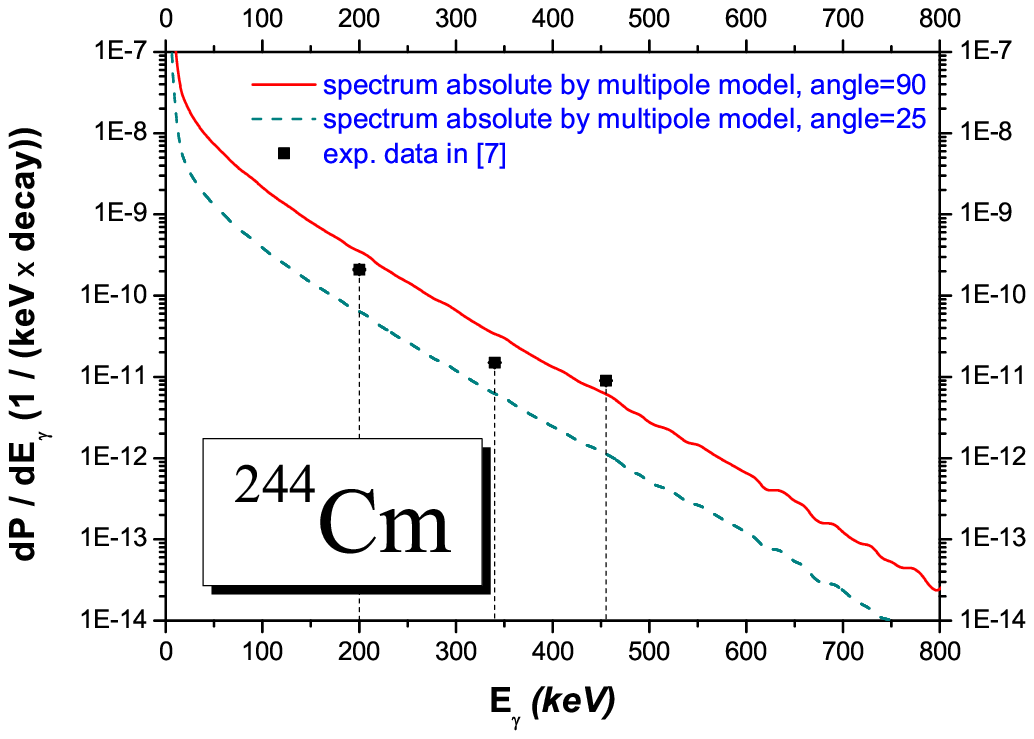}}
\vspace{-5mm}
\caption{The bremsstrahlung probability in the $\alpha$-decay of the $^{214}{\rm Po}$, $^{226}{\rm Ra}$ and $^{244}{\rm Cm}$ nuclei:
red solid line is for the absolute probability calculated by the multipole model,
dash-dot blue line for the normalized spectrum calculated by approach~\cite{Maydanyuk.2008.MPLA},
dash green line for the absolute probability calculated by formula (\ref{eq.2.11.1}) of the multipole model with taking the spectroscopic factor $S_{\alpha}^{\rm (th)}$ into account for the deformed nucleus $^{226}{\rm Ra}$ (we use approximated value $S_{\alpha}^{\rm (th)}=0.2$ from table I in~\cite{Peltonen.2008.PRC}),
scatter line for experimental data
(data~\cite{Maydanyuk.2008.EPJA} for $^{214}{\rm Po}$,
data~\cite{Maydanyuk.2008.MPLA} for $^{226}{\rm Ra}$),
data~\cite{Kasagi.1997.JPHGB,Kasagi.1997.PRLTA} for $^{244}{\rm Cm}$)).
\label{fig.3}}
\end{figure}

The best result in agreement between theory and experiment I have obtained for the $^{214}\mbox{\rm Po}$ nucleus (see the left panel in Fig.~\ref{fig.3}; here there is no any normalization of the calculated curve relatively experimental data). From figure one can see that for this nucleus the calculated spectrum by the proposed approach is in enough good agreement with the experimental data~\cite{Maydanyuk.2008.EPJA} inside the region from 100 keV up to 750 keV.
The calculated absolute probabilities of the bremsstrahlung in $\alpha$-decay of the $^{226}\mbox{\rm Ra}$ nucleus and experimental data in~\cite{Maydanyuk.2008.MPLA} for this nucleus are presented in the central panel in Fig.~\ref{fig.3}. For this nucleus at low energies of the photons emitted the calculated spectra are located below experimental data, but for energies from 350 keV and higher I have obtained good agreement between theory and experiment.
%
%
%
I add the calculated absolute probabilities for the $^{244}{\rm Cm}$ nucleus, comparing them with the high limit of errors of experimental data~\cite{Kasagi.1997.JPHGB,Kasagi.1997.PRLTA}
(see the right panel in Fig.~\ref{fig.3}).
I calculate the bremsstrahlung spectrum for $\vartheta_{\alpha\gamma}=25^{\circ}$ of the angle $\vartheta_{\alpha\gamma}$ between direction of motion of the $\alpha$-particle and photon emission which correspond to experiment~\cite{Kasagi.1997.PRLTA},
and I add another spectrum obtained for $\vartheta_{\alpha\gamma}=90^{\circ}$.
From this figure I see that both calculated curves are located enough close to the high limit of error of experimental data, and perhaps one can conclude that the agreement between theory and experiment is not so bad.
This figure clearly demonstrates that the difference between higher and lower spectra could be explained by different values of this angle!
Such explanation of two different spectra on the basis of one model is given for the first time, and one can suppose that by such a way a question discussed in \cite{Eremin.2000.PRLTA,Kasagi.2000.PRLTA} concerning to the $^{210}{\rm Po}$ nucleus is explained also.


\subsection{Bremsstrahlung dependence on $Q_{\alpha}$ and predictions of the bremsstrahlung probability during $\alpha$-decay of isotopes of ${\rm Th}$
\label{sec.3.4}}

\begin{figure}[htbp]
\centerline{\includegraphics[width=88mm]{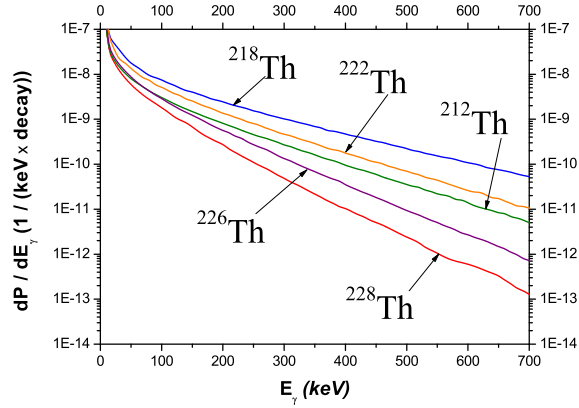}}
\vspace{-5mm}
\caption{\small
The predicted absolute bremsstrahlung probabilities in $\alpha$-decay of the $^{228}{\rm Th}$ nucleus and its isotopes
\label{fig.4}}
\end{figure}

In Ref.~\cite{Ohtsuki.2006.CzJP} it was reported about current investigations of bremsstrahlung accompanying the $\alpha$-decay of the $^{228}{\rm Th}$ nucleus. Let us estimate the absolute bremsstrahlung probability for this nucleus on the basis of the proposed model. Results of such calculations are presented in Fig.~\ref{fig.4}.
In calculations I use~\cite{Buck.1993.ADNDT}:
the angle $\vartheta$ between the directions of the $\alpha$-particle motion (with possible tunneling) and the photon emission is $90^{\circ}$,
$Q_{\alpha}$-value is 5.555~MeV.
In Ref.~\cite{Maydanyuk.2008.MPLA} we explained the difference between the photon emission probabilities (both experimental and theoretical results) in the $\alpha$-decay of $^{226}{\rm Ra}$ and $^{214}{\rm Po}$ (at first, dependence of the bremsstrahlung probability on the $\alpha$-particle energy was analyzed in Ref.~\cite{So_Kim.2000.JKPS}):
\emph{``The difference between the two sets of data can be attributed to the different structure of the two nuclei, which affects the motion of the $\alpha$-particle inside the barrier. The ratio between the two sets of data of the photon emission probability $dP / dE_{\gamma}$ is strongly characterized by the different $\alpha$-decay energy for $^{214}{\rm Po}$ (E$_{\alpha}$=7.7 MeV)  and $^{226}{\rm Ra}$ (E$_{\alpha}$=4.8 MeV) concerning the shapes of the alpha-nucleus barriers for these nuclei.''}
The difference between the $\alpha$-particle energies for the decaying $^{214}{\rm Po}$ and $^{226}{\rm Ra}$ nuclei is directly connected with different tunneling regions for these nuclei, which is directly connected with different contributions of the photons emission from tunneling and external regions, interference terms into the total spectra. And we obtained the property: \emph{The tunneling region is larger, the bremsstrahlung spectrum is smaller.}
The smaller values of the calculated total emission probability for $^{226}{\rm Ra}$ than the one for $^{214}{\rm Po}$ can be explained by a consequence of the fact that outside the barrier the Coulomb field (and its derivative respect to $r$) that acts on the $\alpha$-particle in the case of $^{226}{\rm Ra}$ is smaller than in the case of  $^{214}{\rm Po}$ because the external wide region results for the $^{214}{\rm Po}$ nucleus larger than for $^{226}{\rm Ra}$ and therefore the $\gamma$-emission probability for the $^{214}{\rm Po}$ nucleus is bigger.
In Fig.~\ref{fig.4} one can see the demonstration of this property for isotopes of ${\rm Th}$.
In Table~\ref{table.1} it is shown how the bremsstrahlung probability depends on $Q_{\alpha}$-value of the nucleus for different energies of the photons emitted.
\begin{table}
\begin{center}
\begin{tabular}{|c|c|c|c|c|c|c|c|c|} \hline
 \multicolumn{4}{|c|}{$\alpha$-decay data} &
 \multicolumn{5}{|c|}{Bremsstrahlung probability, 1 / keV / decay}
 \\ \cline{1-9}
  $A_{p}$ &
  $Q_{\alpha}$, MeV &
  $b_{\alpha}^{\rm abs}$, \% &
  $T_{1/2,\, \alpha}^{\rm exp}$, sec &

  100 keV &
  200 keV &
  300 keV &
  400 keV &
  500 keV \\ \hline
  212 & 7.987 & 100.0 & 3.0\,E-2   & 3.0\,E-9 & 8.1\,E-10 & 2.7\,E-10 & 9.5\,E-11 & 3.5\,E-11 \\
  218 & 9.881 & 100.0 & 1.1\,E-7   & 7.5\,E-9 & 2.5\,E-9  & 1.0\,E-9  & 4.7\,E-10 & 2.2\,E-10 \\
  222 & 8.164 & 100.0 & 2.8\,E-3   & 5.2\,E-9 & 1.3\,E-9  & 4.6\,E-10 & 1.7\,E-10 & 7.0\,E-11 \\
  226 & 6.487 & 75.5  & 2.5\,E+3   & 2.9\,E-9 & 5.6\,E-10 & 1.3\,E-10 & 3.5\,E-11 & 9.4\,E-12 \\
  228 & 5.555 & 72.7  & 8.3\,E+7   & 1.8\,E-9 & 2.8\,E-10 & 4.9\,E-11 & 1.0\,E-11 & 1.9\,E-12 \\
  \hline
\end{tabular}
\end{center}
\caption{Estimated values of the bremsstrahlung probability during $\alpha$-decay of the $^{228}{\rm Th}$ nucleus and its isotopes
\label{table.1}}
\end{table}

\subsection{How much is the bremsstrahlung probability changed in dependence on numbers of protons and neutrons of the $\alpha$-decaying nucleus?
\label{sec.3.5}}

Let us analyze how much the probability of photons emitted could change if to change number of protons or neutrons of the nucleus which decays.
\begin{figure}[htbp]
\centerline{%
\includegraphics[width=95mm]{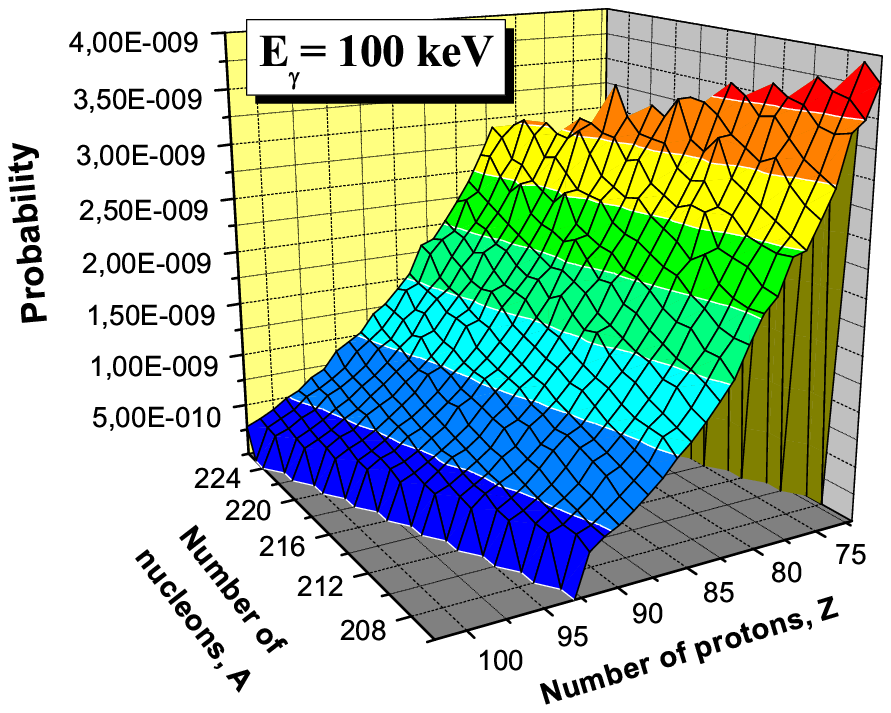}
\hspace{-10mm}\includegraphics[width=95mm]{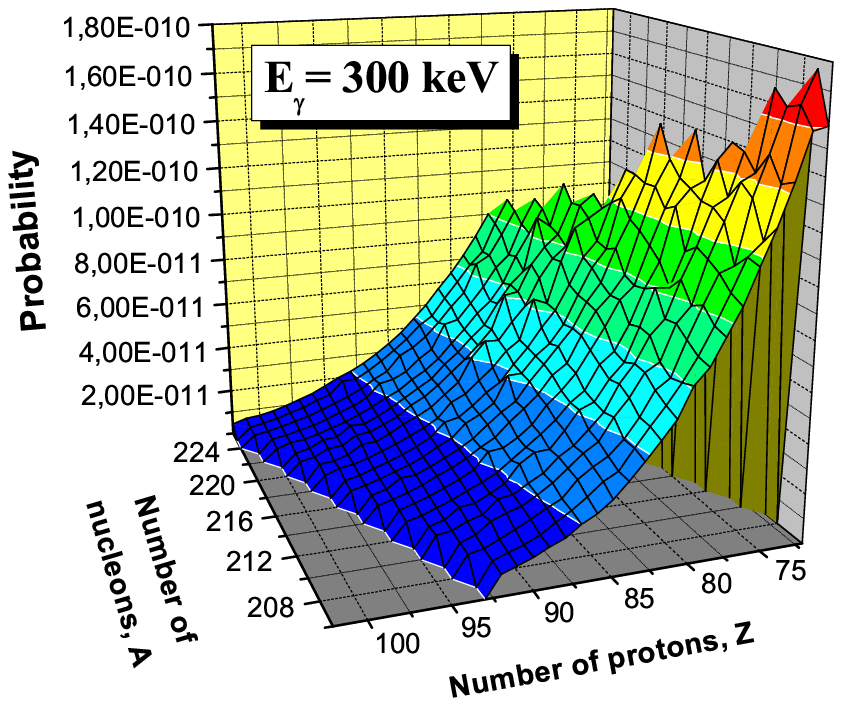}}
\vspace{-4mm}
\caption{Distribution of the absolute bremsstrahlung probability in the $\alpha$-decay on numbers of protons and nucleons of the decaying nucleus (used data: $Q_{\alpha}=5.439$~MeV for $^{210}{\rm Po}$):
left panel is for the probability of the photons emitted with energy $E_{\gamma}=100$~keV,
right panel for the probability of the photons emitted with energy $E_{\gamma}=300$~keV
\label{fig.5}}
\end{figure}
In order to made such analysis clearer, I have fixed Q-value and have calculated the probabilities in some region of numbers of protons and nucleons (it is convenient to use angle between directions of the emission of photon and motion of the $\alpha$-particle equal to $90^{\circ}$, as results are supposed to be similar for any other angles). In the next Fig.~\ref{fig.5} a distribution of the absolute probability in dependence on the numbers of protons and neutrons of the $\alpha$-decaying nucleus is presented
for 100~keV, 300~keV and 500~keV of the emitted photons.
One can see that at increasing of the number of neutrons the probability is changed a little and monotonously while at increasing of the number of protons the probability decreases stronger essentially and monotonously (such proton dependence could be a quantitative demonstration of results of \cite{So_Kim.2000.JKPS}).


\subsection{Angular emission of photons 
\label{sec.3.6}}

\begin{figure}[htbp]
\centerline{%
\includegraphics[width=85mm]{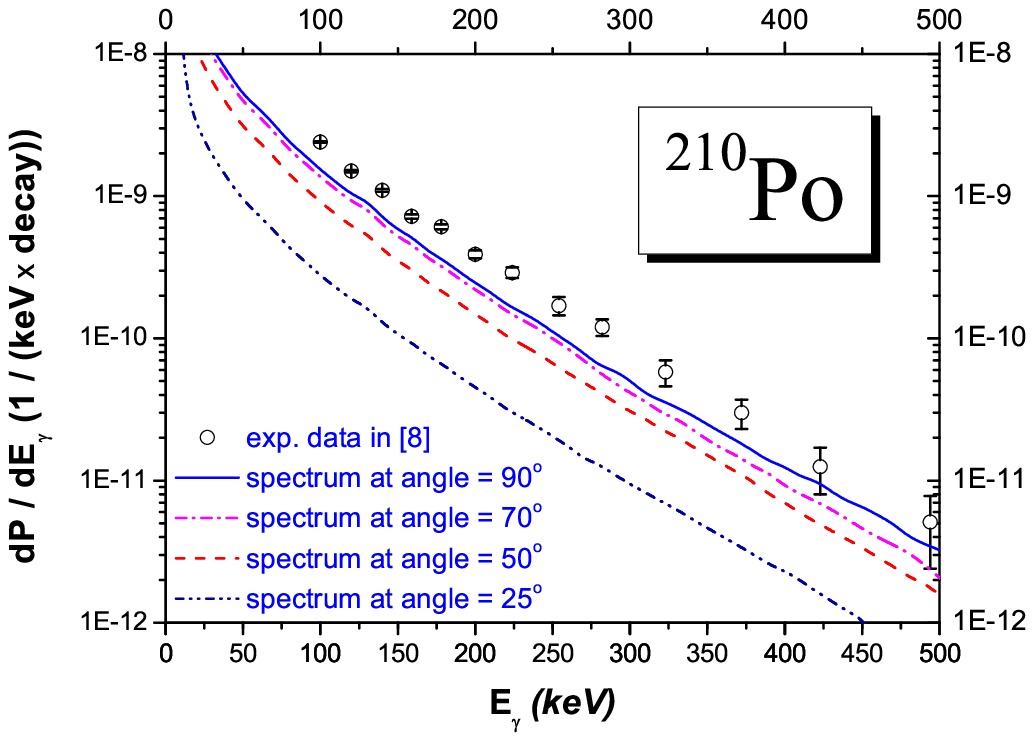}
\includegraphics[width=85mm]{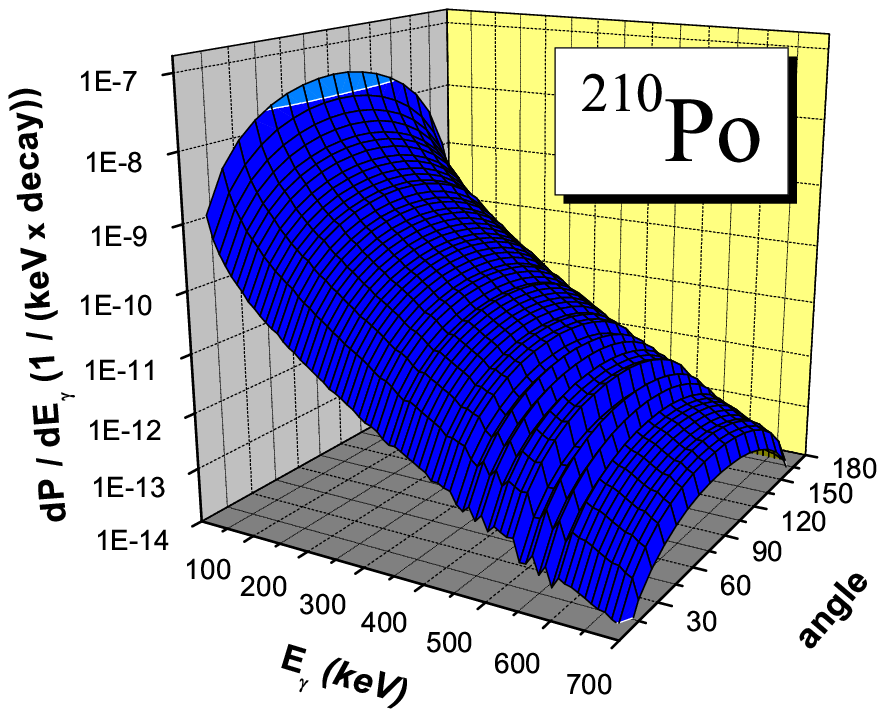}}
\vspace{-4mm}
\caption{The absolute bremsstrahlung probability in the $\alpha$-decay of $^{210}{\rm Po}$ in dependence on the angle $\vartheta_{\alpha\gamma}$ between directions of the motion of the $\alpha$-particle and the emission of photon calculated by the multipole model
\label{fig.6}}
\end{figure}

Now let us analyze how much the bremsstrahlung probability is changed in dependence on the angle between directions of the motion of the particle and emission of photon. For analysis we consider the $^{210}{\rm Po}$ nucleus. In  Fig.~\ref{fig.6} the bremsstrahlung probabilities calculated by the multipole model at different values of this angle
are presented.
In particular, one can see that spectrum for the angle $25^{\circ}$ is located enough far below then the spectrum for the $90^{\circ}$. Now these calculations confirm (at first time) arguments proposed for explanation of difference between experimental data \cite{Kasagi.1997.PRLTA} and \cite{D'Arrigo.1994.PHLTA} obtained for these angles, correspondingly, and they have resolved discussions in \cite{Eremin.2000.PRLTA,Kasagi.2000.PRLTA}.
In the next Fig.~\ref{fig.7} distribution of the bremsstrahlung probability on the angle $\vartheta_{\alpha\gamma}$ between directions of the motion of the particle and emission of photon for the $^{210}{\rm Po}$ nucleus is presented.
\begin{figure}[htbp]
\centerline{%
\includegraphics[width=85mm]{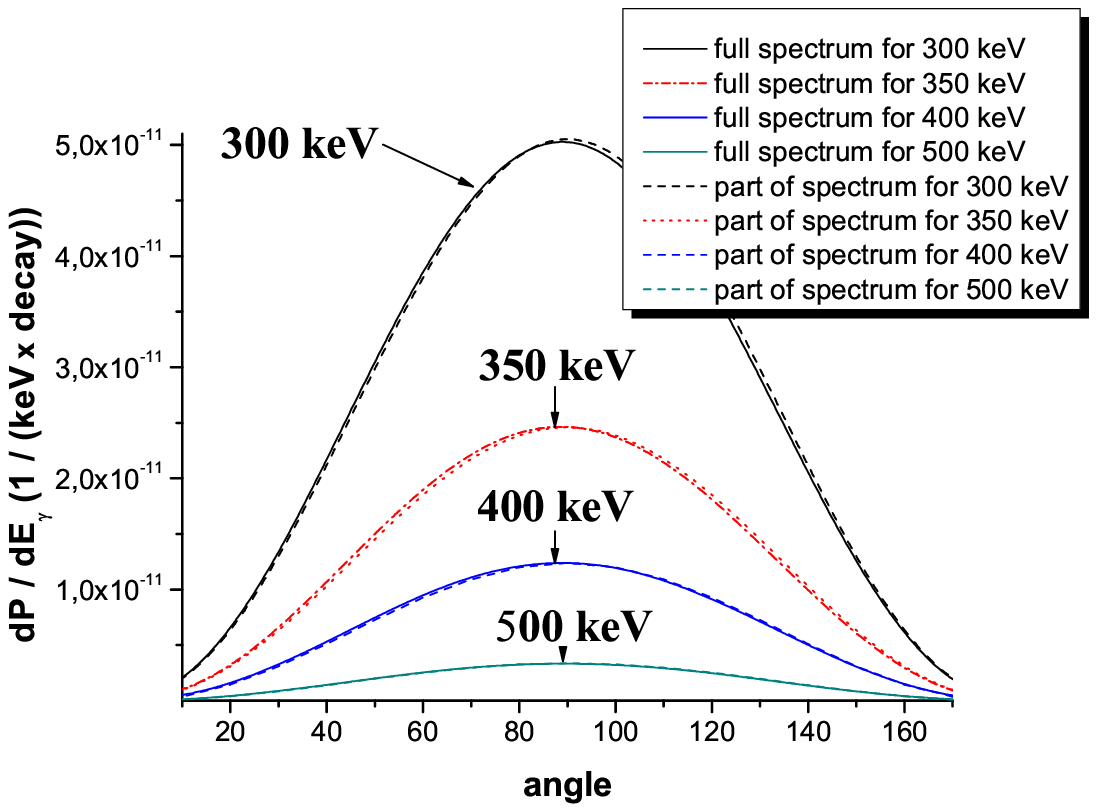}
\includegraphics[width=85mm]{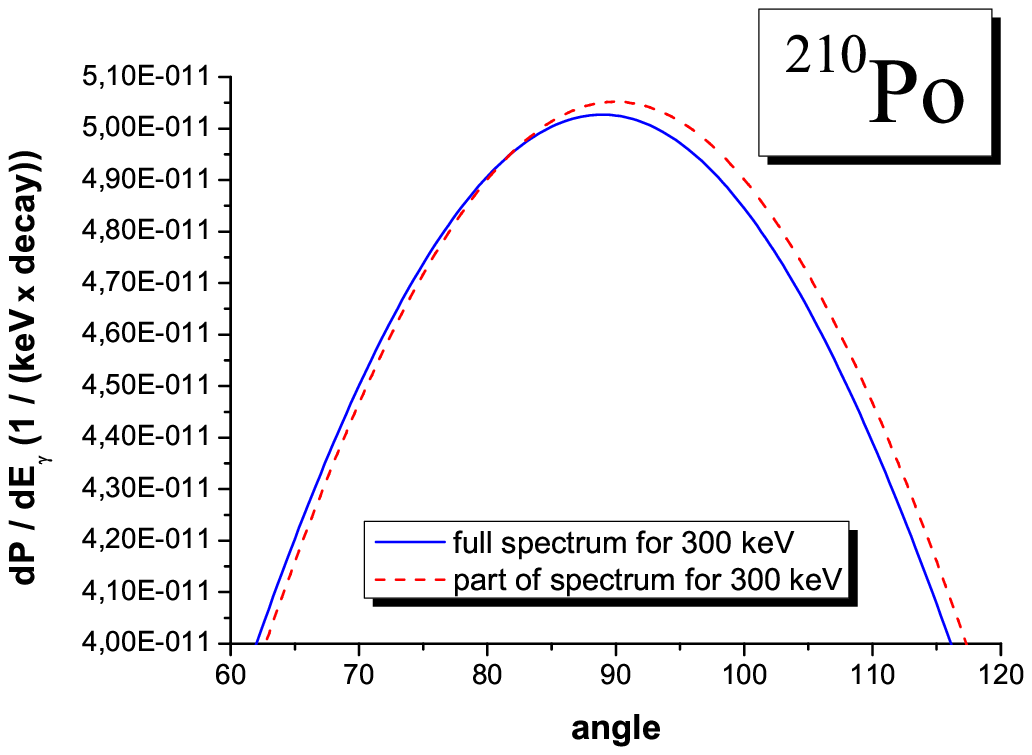}}
\vspace{-4mm}
\caption{Distribution of the bremsstrahlung probability on the angle between directions of the motion of the particle and emission of photon for the $^{210}{\rm Po}$ nucleus:
(a) angular probability at 300~keV, 350~keV, 400~keV, 500~keV;
(b) angular probability at 300~keV in large scale: here one can see difference between two spectra.
\label{fig.7}}
\end{figure}
In particular, one can see that the angular distribution of the probability is formed mainly by the first integral $J(1,0)$ while next two integrals $J(1,1)$ and $J(1,2)$ give very small contribution to the total angular spectrum.


\subsection{Formula of the bremsstrahlung probability in the $\alpha$-decay
\label{sec.3.7}}


Let us restrict ourselves by only one nucleus and try to write such formula for it.
After preliminary estimations of the spectra for different nuclei, I propose the following form
(further, let's use only $\vartheta_{\alpha\gamma}=90^{\circ}$):
\begin{equation}
\begin{array}{ccl}
  \ln \;
  \biggl\{
    \displaystyle\frac
      {d P_{\rm param}(w;\, a_{0}\ldots a_{4},\, n_{1}\ldots n_{4})}
      {d\,\Omega_{ph} \: d\cos{\theta_{f}}}
  \biggr\} & = &
  \ln \;
  \biggl\{
    \displaystyle\frac{e^{2}}{8\,\pi\, c^{5}}\:
    \displaystyle\frac{Z_{eff}^{2}\, E_{i}}{m^{2}\, k_{i}}
  \biggr\} +
  a_{0} -
  a_{1}\,w^{n_{1}} +
  \displaystyle\frac{a_{2}}{w^{n_{2}}} +
  \displaystyle\frac{a_{3}}{w^{n_{3}}} +
  \displaystyle\frac{a_{4}}{w^{n_{4}}},
\end{array}
\label{eq.3.3.1}
\end{equation}
where $a_{0}\ldots a_{4}$ and $n_{1}\ldots n_{4}$ are unknown constants which do not depend on energy of the photon emitted and are changed for the different nuclei. These constants reflect ``structure'' of the $\alpha$-decay for the studied nucleus. Therefore, they should depend on $Q_{\alpha}$, $Z_{\rm eff}$, $Z_{d}$, $A_{d}$ of this nucleus.

With a purpose to find parameters $a_{0}\ldots a_{4}$ $n_{1}\ldots n_{4}$ for the selected nucleus, I shall introduce the following characteristic:
\begin{equation}
\begin{array}{ccl}
  \vspace{2mm}
  \sigma\, (a_{i},\, n_{i}) & = &
  \sqrt{
    \displaystyle\frac{1}{w_{\rm max} - w_{\rm min}}\;
    \displaystyle\int\limits_{w_{\rm min}}^{w_{\rm max}}
    \Bigl[\Delta P\;(w;\; a_{i},\, n_{i})\Bigr]^{2} \; dw}, \\

  \Delta P\; (w;\; a_{i},\, n_{i}) & = &
  \ln\;\biggl\{ \displaystyle\frac{d P_{\rm model}(w)} {d\,\Omega_{ph} \: d\cos{\theta_{f}}} \biggr\}\; - \;
  \ln\;\biggl\{ \displaystyle\frac{d P_{\rm param} (w;\, a_{0}\ldots a_{4},\, n_{1}\ldots n_{4})}
                                  {d\,\Omega_{ph} \: d\cos{\theta_{f}}} \biggr\},
\end{array}
\label{eq.3.3.2}
\end{equation}
where $d P_{\rm model}$ and $d P_{\rm param}$ are the bremsstrahlung probabilities calculated by the multipole model and by the formula (\ref{eq.3.3.1}), correspondingly. The $\sigma$ at selected set of parameters is smaller, the curve $d P_{\rm param}$ obtained by formula (\ref{eq.3.3.1}) is closer to the spectrum $d P_{\rm model}$ calculated by the multipole model. I.~e. the best description of the bremsstrahlung spectrum for the studied nucleus by formula (\ref{eq.3.3.1}) should be obtained at such choice of the parameters $a_{0}\ldots a_{4}$ and $n_{1}\ldots n_{4}$ where $\sigma$ is minimal. For convenience, I call this approach for determination of parameters for the selected nucleus as \emph{method of minimization}.
So, using the method of minimization, for the $^{218}{\rm Th}$ nucleus I obtain the following values ($w_{\rm min}=50$~keV and $w_{\rm max}=900$~keV are used):
\begin{equation}
\begin{array}{ccccc}
  n_{1} = 1, &
  n_{2} = 0.5, &
  n_{3} = 1, &
  n_{4} = 2, & \\

  a_{0} = 10.8, &
  a_{1} = 0.007, &
  a_{2} = 10, &
  a_{3} = 10, &
  a_{4} = 1.
\end{array}
\label{eq.3.3.3}
\end{equation}
The curve calculated by formula~(\ref{eq.3.3.1}) at choice (\ref{eq.3.3.3}) of parameters turns out to be located extremely close to the bremsstrahlung spectrum for $^{218}{\rm Th}$! From here I conclude that \emph{the bremsstrahlung probability for arbitrary one nucleus can be approximated by formula (\ref{eq.3.3.1}) with very high accuracy inside the energy region up to 1~MeV.} Estimations of parameters for other nuclei show that it is possible to describe the bremsstrahlung spectra with enough high accuracy for different nuclei using different values of the $a_{0}$ and $a_{1}$ parameters only, while the $n_{1} \ldots n_{4}$ parameters and even the $a_{2}$, $a_{3}$, $a_{4}$ parameters are fixed.
%

In this paper, I shall define the $n_{1}\ldots n_{4}$, $a_{2}$, $a_{3}$ and $a_{4}$ parameters for the different nuclei by (\ref{eq.3.3.3}) and try to find out how $a_{0}$ and $a_{1}$ can be described. Assuming dependence of $a_{0}$ and $a_{1}$ on $Q$, $A_{d}$ and $Z_{d}$ to be linear,
I propose the following formula:
\begin{equation}
\begin{array}{ccl}
  a_{0}\, (Q, A_{d}, Z_{d}) & = & b_{00} + b_{01}\,Q + b_{02}\,A_{d} + b_{03}\,Z_{d}, \\
  a_{1}\, (Q, A_{d}, Z_{d}) & = & b_{10} + b_{11}\,Q + b_{12}\,A_{d} + b_{13}\,Z_{d},
\end{array}
\label{eq.3.3.4}
\end{equation}
where new unknown parameters $b_{0i}$ and $b_{1i}$ ($i=0,1,2,3$) have been introduced which do not already depend on $Q$, $Z_{d}$ and $A_{d}$.
The simplest way is to find $b_{01}$ and $b_{11}$. For such calculations we needs in two nuclei with equal $Z_{d}$ and $A_{d}$ values and different $Q$-values. Let's consider the $^{228}{\rm Th}$ nucleus. For it I calculate the bremsstrahlung spectrum on the basis of the multipole model at two different $Q$-values (I use: $Q_{1}=5.555$~MeV and $Q_{2}=10$~MeV), and then I obtain the $a_{0}$ and $a_{1}$ parameters using the method of minimization above.
Results are presented in Table~\ref{table.2} in the first two strings with numbers 1 and 2.
\begin{table}
\begin{center}
\begin{tabular}{|c|c|c|c|c|c|c|c|c|c|} \hline
 \multicolumn{6}{|c|}{$\alpha$-decay data} &
 \multicolumn{4}{|c|}{parameters}
 \\ \cline{1-10}
  No. &
  $A_{d}$ &
  $A_{d}^{1/3}$ &
  $Z_{d}$ &
  $Z_{\rm eff}$ &
  $Q_{\alpha}$, MeV &
  $a_{0}^{\rm (min)}$ & $a_{0}^{\rm (param)}$ & $a_{1}^{\rm (min)}$ & $a_{1}^{\rm (param)}$
  \\ \hline
  1 & 224 & 6.073177 &  88 & 0.42105 & 5.555 &   10.2 & 10.20083 & 0.0154  & 0.01531749 \\
  2 & 224 & 6.073177 &  88 & 0.42105 & 10.0  &   11.2 & 11.20020 & 0.0069  & 0.00681732 \\
  3 & 102 & 4.672328 &  50 & 0.03774 & 10.0  &    6.3 &  6.30084 & 0.00475 & 0.00440210 \\
  4 & 262 & 6.398827 & 107 & 0.36090 & 10.0  &   10.9 & 10.90000 & 0.008   & 0.00799993 \\ \hline
\end{tabular}
\end{center}
\caption{Parameters $a_{0}$ and $a_{1}$ for $^{228}{\rm Th}$, $^{106}{\rm Te}$ and nucleus with $A_{p}=266$ and $Z_{p}=109$
($a_{0}^{\rm (min)}$ and $a_{1}^{\rm (min)}$ are parameters calculated by the method of minimization,
$a_{0}^{\rm(param)}$ and $a_{1}^{\rm(param)}$ are parameters calculated by formula (\ref{eq.3.3.9}))
\label{table.2}}
\end{table}

\begin{figure}[htbp]
\centerline{%
\includegraphics[width=60mm]{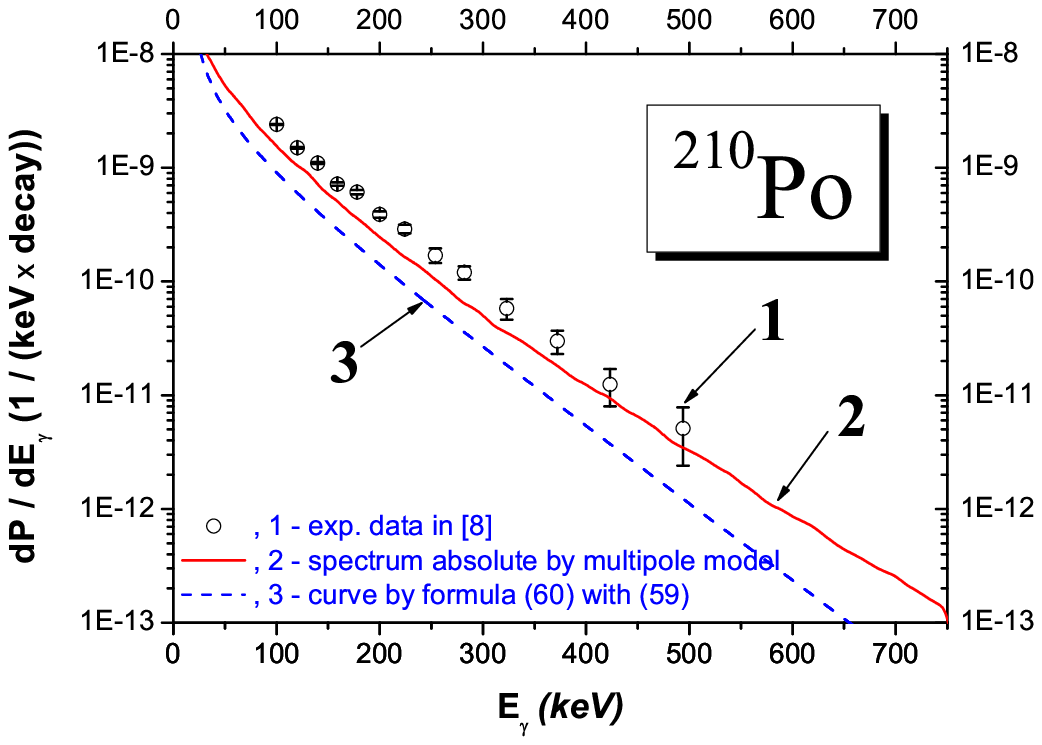}
\includegraphics[width=60mm]{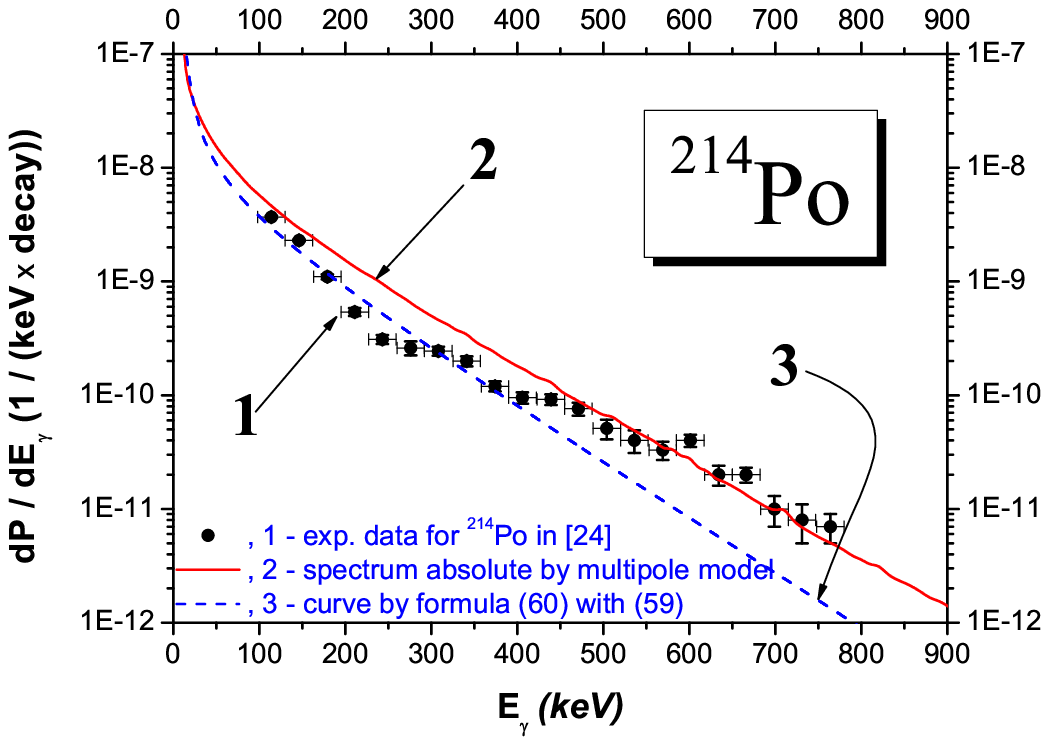}
\includegraphics[width=60mm]{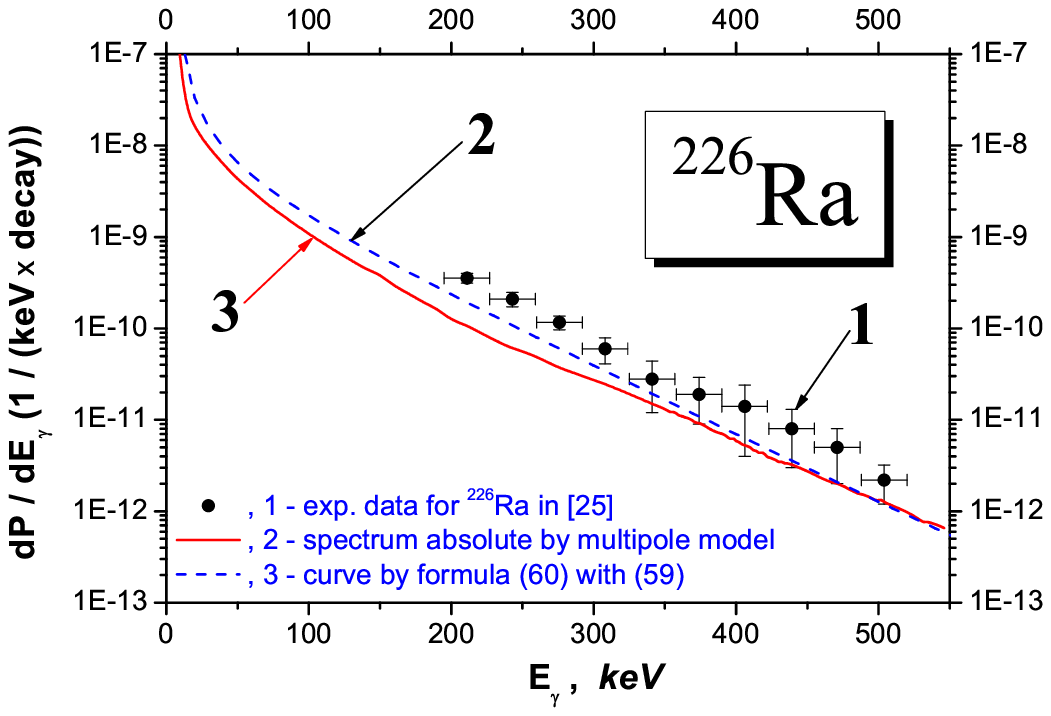}}
\vspace{-6mm}
\caption{The bremsstrahlung probability in the $\alpha$-decay of the $^{210}{\rm Po}$, $^{214}{\rm Po}$ and $^{226}{\rm Ra}$ nuclei ($\vartheta_{\alpha\gamma}=90^{\circ}$): the absolute probability calculated by the multipole model (red solid line), experimental data (scatter line, data~\cite{Maydanyuk.2008.EPJA} for $^{214}{\rm Po}$, data~\cite{Boie.2007.PRL} for $^{210}{\rm Po}$ and data~\cite{Maydanyuk.2008.MPLA} for $^{226}{\rm Ra}$)
and curve calculated by formula (\ref{eq.3.3.1}) with (\ref{eq.3.3.9}) (dash blue line).
\label{fig.8}}
\end{figure}

In order to find next unknown parameters $b_{ij}$ ($i=1,2$, $j= 0,2,3$), it needs to consider nuclei with the different $A_{d}$, $Z_{d}$ numbers at the same $Q$-value. To achieve accuracy as high as possible, we shall use the previous nucleus and two other nuclei with the largest difference between $A_{d}$ (and between $Z_{d}$). Let's use Table in \cite{Buck.1993.ADNDT}, from here we select: $^{106}{\rm Te}$ and nucleus with $A_{p}=266$, $Z_{p}=109$.
I calculate the bremsstrahlung spectra at $Q_{\alpha}$-value equals to 10~MeV using the multipole model and then I find $a_{0}$ and $a_{1}$ for them using the minimization method. Results are presented in Table~\ref{table.2} in the next two strings with numbers 3 and 4. Using data of the table~\ref{table.2}, I calculate unknown $b_{0i}$ and $b_{1i}$, then I find the final form of $a_{0}$ and $a_{1}$ on $Q$, $A_{d}$ and $Z_{d}$
($Q$ is used in MeV):
\begin{equation}
\begin{array}{ccl}
  a_{0}\,(Q, A_{d}, Z_{d}) & = &
  4.60202 + 0.22497 \cdot Q + 0.11956 \cdot A_{\rm d} - 0.25492 \cdot Z_{\rm d}, \\
  a_{1}\,(Q, A_{d}, Z_{d}) & = &
  0.0204108 - 0.0019123 \cdot Q + 1.086956\cdot 10^{-6} \cdot A_{\rm d} + 6.0068649\cdot 10^{-5} \cdot Z_{\rm d}
\end{array}
\label{eq.3.3.9}
\end{equation}
and the bremsstrahlung formula (\ref{eq.3.3.1}) has transformed into such:
\begin{equation}
\begin{array}{ccl}
  \ln \;
  \biggl\{
    \displaystyle\frac{d P^{E1+M1}_{1}(w,\, \theta_{f}=90^{\circ})} {d\,\Omega_{ph} \: d\cos{\theta_{f}}}
  \biggr\} & = &
  \ln \;
  \biggl\{
    \displaystyle\frac{e^{2}}{8\,\pi\, c^{5}}\:
    \displaystyle\frac{Z_{eff}^{2}\, E_{i}}{m^{2}\, k_{i}}
  \biggr\} +
  a_{0} -
  a_{1}\,w +
  \displaystyle\frac{10}{\sqrt{w}} +
  \displaystyle\frac{10}{w} +
  \displaystyle\frac{1}{w^{2}}.
\end{array}
\label{eq.3.3.10}
\end{equation}
For four studied nuclei I have obtained very small difference between the probability calculated by the model above and the curve calculated by formula (\ref{eq.3.3.10}) with parameters (\ref{eq.3.3.9}) for the energy of photons emitted up to 1 MeV (less then 1 percent). \emph{I.~e. we have described the bremsstrahlung spectra inside the energy region up to 1 MeV for four different nuclei (with such long maximal distance between their numbers $A_{d}$) with very good accuracy by only one this formula with parameters calculated only on the basis of values $A_{d}$, $Z_{d}$, $Q_{\alpha}$!} It turns out that description of the bremsstrahlung spectra for all nuclei inside region $A_{d}= 107..262$ at different $Z_{d}$ by the formula (\ref{eq.3.3.10}) with parameters (\ref{eq.3.3.9}) is not such accurate but enough good also (see curves in Fig.~\ref{fig.8} obtained by formula). However, one can improve further such approximation essentially for the ``problem'' nuclei, if to pass from the linear dependence (\ref{eq.3.3.4}) of the bremsstrahlung probability on the $A_{d}$ and $Z_{d}$ values to harmonic one.


\section{Conclusions}

In the paper the model of the bremsstrahlung accompanying the $\alpha$-decay is presented where emphasis is given on development of unified angular formalism of the dipole and multipole approaches. Effectiveness of the model and accuracy of calculations of the bremsstrahlung spectra are analyzed in their comparison with experimental data for the $^{210}{\rm Po}$, $^{214}{\rm Po}$, $^{226}{\rm Ra}$ and $^{244}{\rm Cm}$ nuclei.
Note the following.
\begin{itemize}
\item
The multipolar model is the most motivated from the physical point of view, it is the richest in obtaining useful information about angular emission of photons during $\alpha$-decay, their results obtained for the $^{210}\mbox{\rm Po}$, $^{214}\mbox{\rm Po}$ and $^{226}\mbox{\rm Ra}$ nuclei both in the absolute scale and with normalization on experimental data \cite{Boie.2007.PRL,Maydanyuk.2008.EPJA,Maydanyuk.2008.MPLA} are in the best agreement with these experimental data in comparison with any other known models and approaches.

\begin{itemize}
\item
The best result has been obtained in agreement between the calculated absolute probability of the bremsstrahlung emission for the $^{214}\mbox{\rm Po}$ nucleus and the experimental data\cite{Maydanyuk.2008.EPJA} for this nucleus inside the region of photons energies from 100 keV up to 750 keV (see Fig.~\ref{fig.3} (a), $Q_{\alpha} = 7.865$~MeV, the angle $\vartheta_{\alpha\gamma}$ between the directions of the $\alpha$-particle motion and the photon emission is $90^{\circ}$).

\item
The calculated absolute probabilities of the bremsstrahlung emission in $\alpha$-decay of the $^{210}\mbox{\rm Po}$ and $^{226}\mbox{\rm Ra}$ nuclei for low energies of the photons emitted are located below experimental data \cite{Boie.2007.PRL} and \cite{Maydanyuk.2008.MPLA}, but for energies from 350 keV and higher I have obtained good agreement between the model and experiment (see Fig.~\ref{fig.3} (b) and (c), $Q_{\alpha} = 5.439$~MeV for $^{210}\mbox{\rm Po}$ and $Q_{\alpha} = 4.904$~MeV for $^{226}\mbox{\rm Ra}$, $\vartheta_{\alpha\gamma}=90^{\circ}$).

\item
The spectrum for the $^{244}{\rm Cm}$ nucleus obtained at $\vartheta_{\alpha\gamma}=25^{\circ}$ is found to be in satisfactory agreement with high limit of errors of experimental data \cite{Kasagi.1997.PRLTA}.

\end{itemize}
\item
In frameworks of the dipole approach, the emission of photons from the internal region before the barrier is not small and gives visible contribution into the total spectrum. However, the multipolar approach seems to be less sensitive to such emission.
Such a peculiarity could be explained by more accurate use of the far asymptotic region by the multipole approach, while the dipole approach reduces this region and, so, has higher convergence in numerical integration of the spectra over $r$. However, more accurate use of the far asymptotic region by the multipole approach provides essentially better agreement between obtained spectra and experimental data, while the dipole approach seems to be fail here (at its present development). This important point adds the power of predictions in absolute scale to the multipolar model, in contrast to the dipole one.
This demolishes published progress in agreement between experimental data and the spectra calculated in the dipole approach (where such emission from the internal region was neglected) if it was affirmed as obtained in the absolute scale.


\item
A hypothesis about explanation of difference between two experiments \cite{D'Arrigo.1994.PHLTA} and \cite{Kasagi.1997.JPHGB} on the basis of different values $25^{\circ}$ and $90^{\circ}$ of the angle $\vartheta_{\alpha\gamma}$ between direction of motion of the $\alpha$-particle and emission of photons proposed in discussions \cite{Eremin.2000.PRLTA,Kasagi.2000.PRLTA}, has been confirmed (at the first time).

\item
The unified formula of the bremsstrahlung probability (in the absolute scale) during the $\alpha$-decay of the arbitrary nucleus, which is directly expressed through the $Q_{\alpha}$-value and numbers $A_{d}$, $Z_{d}$ of protons and neutrons of this nucleus, has been constructed. Inside region of the $\alpha$-active nuclei from $^{106}{\rm Te}$ up to the nucleus with numbers of nucleons and protons $A_{p}=266$ and $Z_{p}=109$ (this region is taken from \cite{Buck.1993.ADNDT}) with energy of the photons emitted from 50 keV up to 900 keV satisfactory agreement has been achieved between the spectra, obtained on the basis of the multipole model (where duration of calculations for one selected nucleus is up to 1 day), and the bremsstrahlung spectra obtained on the basis of the proposed formula (where duration of calculations is about some seconds using the same computer). However, analyzing results for the $^{210}\mbox{\rm Po}$, $^{210}\mbox{\rm Po}$ and $^{226}\mbox{\rm Ra}$ nuclei, this formula is found to give more accurate spectra relatively experimental data (up to 500~keV), then spectra obtained in the dipole approach without taking the emission from the internal region before the barrier into account.
\end{itemize}
%


\section*{Acknowledgements
\label{sec.acknowledgements}}

The author is appreciated to
Dr. Alexander~K.~Zaichenko for his assistance in computer realization of numerical methods in calculations of wave functions,
Prof.~Vladislav~S.~Olkhovsky for useful discussions concerning realizations of multiple internal reflections in the problem of $\alpha$-decay and comments of definition of phase times,
Prof.~Giorgio~Giardina for useful discussions concerning main formalism of the model, dependence of the bremsstrahlung spectra on $Q_{\alpha}$-value of $\alpha$-decay, aspects to investigate deformed nuclei in this problem,
Prof.~Volodimir~M.~Kolomietz for useful discussions and critical comments concerning the general formalism of the presented model,
Dr.~Sergei~N.~Fedotkin for useful comments concerning definitions of absolute and normalized probabilities of the photons emitted during the $\alpha$-decay,
Dr.~Alexander~G.~Magner for useful comments concerning determination of wave function of the $\alpha$-decaying system,
Dr.~Vladislav Kobychev for interesting discussions with some analysis concerning behavior of the bremsstrahlung spectra for photon energies close to zero,
Dr.~Volodymyr Kyrytsya for useful comments how to improve the computer integration of matrix elements at large radius $r$ (in asymptotic region) with increasing of convergence and keeping of accuracy in calculations.


\appendix
\section{Clebsch-Gordan coefficients
\label{app.3}}

\vspace{5mm}
I define Clebsch-Gordan coefficients, according to Table~I in ref.\cite{Eisenberg.1973} (see~p.317) and find:
\begin{equation}
\begin{array}{lll}
  (0 1 1 \big| \,2,  -1, \,1) = 0, &
  (1 1 1 \big| \,2,  -1, \,1) = 0, &
  (2 1 1 \big| \,2,  -1, \,1) = \sqrt{\displaystyle\frac{3}{5}}, \\

  (0 1 1 \big| \,0, \,1, \,1) = \sqrt{\displaystyle\frac{1}{2}}, &
  (1 1 1 \big| \,0, \,1, \,1) = -\sqrt{\displaystyle\frac{1}{2}}, &
  (2 1 1 \big| \,0, \,1, \,1) = \sqrt{\displaystyle\frac{1}{10}}, \\

  (0 1 1 \big| \,0,  -1, -1)  = \sqrt{\displaystyle\frac{1}{2}}, &
  (1 1 1 \big| \,0,  -1, -1)  = \sqrt{\displaystyle\frac{1}{2}}, &
  (2 1 1 \big| \,0,  -1, -1)  = \sqrt{\displaystyle\frac{1}{10}}, \\

  (0 1 1 \big|  -2, \,1, -1)  = 0, &
  (1 1 1 \big|  -2, \,1, -1)  = 0, &
  (2 1 1 \big|  -2, \,1, -1)  = \sqrt{\displaystyle\frac{3}{5}}.
\end{array}
\label{eq.app.3.1}
\end{equation}

\section{Coefficients $C_{l_{f} l_{ph} n}^{m \mu^{\prime}}$
\label{app.4}}

\vspace{5mm}
We define the coefficient $C_{l_{f} l_{ph} n}^{m \mu^{\prime}}$ so:
\begin{equation}
\begin{array}{lcl}
  C_{l_{f} l_{ph} n}^{m \mu^{\prime}} & = &
    (-1)^{l_{f}+n+1 - \mu^{\prime} + \frac{|m+\mu^{\prime}|}{2}} \;
    (n, 1, l_{ph} \big| -m-\mu^{\prime}, \mu^{\prime}, -m)\; \times \\

  & & \times\;
    \sqrt{\displaystyle\frac{(2l_{f}+1)\,(2n+1)}{32\pi}\;
          \displaystyle\frac{(l_{f}-1)!}{(l_{f}+1)!} \;
          \displaystyle\frac{(n-|m+\mu^{\prime}|)!}{(n+|m+\mu^{\prime}|)!}}
\end{array}
\label{eq.app.4.1}
\end{equation}
At $l_{f}=1$, $l_{ph}=1$ and $n=0$ we have:
\begin{equation}
  m = - \mu^{\prime} = \pm 1
\label{eq.app.4.2}
\end{equation}
and the coefficient $C_{l_{f} l_{ph} n}^{m \mu^{\prime}}$ is:
\begin{equation}
\begin{array}{lcl}
  C_{110}^{m \mu^{\prime}} & = &
    - \sqrt{\displaystyle\frac{3}{64\pi}} \cdot (0 1 1 \big| \; 0, \mu^{\prime}, -m).
\end{array}
\label{eq.app.4.3}
\end{equation}
At $l_{f}=1$, $l_{ph}=1$ and $n=1$ the property (\ref{eq.app.4.2}) is fulfilled and we obtain:
\begin{equation}
\begin{array}{lcl}
  C_{111}^{m \mu^{\prime}} & = &
    \sqrt{\displaystyle\frac{9}{64\pi}} \cdot (1 1 1 \big| \; 0, \mu^{\prime}, -m).
\end{array}
\label{eq.app.4.4}
\end{equation}
At $l_{f}=1$, $l_{ph}=1$ and $n=2$ the property (\ref{eq.app.4.2}) is not fulfilled and
\begin{equation}
\begin{array}{lcl}
  C_{112}^{m \mu^{\prime}} & = &
    (-1)^{- \mu^{\prime} + \frac{|m+\mu^{\prime}|}{2}} \;
    \sqrt{\displaystyle\frac{15}{64\,\pi}\;
          \displaystyle\frac{(2-|m+\mu^{\prime}|)!}{(2+|m+\mu^{\prime}|)!}} \cdot
    (2 1 1 \big| -m-\mu^{\prime}, \mu^{\prime}, -m).
\end{array}
\label{eq.app.4.5}
\end{equation}
Rewrite these coefficients at different $m = \pm 1$ and $\mu^{\prime} = \pm 1$:
\begin{equation}
\begin{array}{lcllcl}
  C_{112}^{-1 -1} & = &
    \displaystyle\frac{1}{16} \: \sqrt{\displaystyle\frac{5}{2\,\pi}} \cdot (2 1 1 \big| \;2, -1, 1), &

  C_{112}^{-1 1} & = &
    - \displaystyle\frac{1}{8} \sqrt{\displaystyle\frac{15}{\pi}} \cdot (2 1 1 \big| \;0 1 1), \\

  C_{112}^{1 -1} & = &
    - \displaystyle\frac{1}{8} \sqrt{\displaystyle\frac{15}{\pi}} \cdot (2 1 1 \big| \;0, -1, -1), &

  C_{112}^{11} & = &
    \displaystyle\frac{1}{16} \: \sqrt{\displaystyle\frac{5}{2\,\pi}} \cdot (2 1 1 \big| \:-2, 1, -1).
\end{array}
\label{eq.app.4.6}
\end{equation}
Substituting here values (\ref{eq.app.3.1}) for the Clebsh-Gordan coefficients, we find:
\begin{equation}
\begin{array}{llll}
  \vspace{2mm}
  C_{110}^{-1 -1} = 0, &
  C_{110}^{-1 1} = -\displaystyle\frac{1}{8} \cdot \sqrt{\displaystyle\frac{3}{2\,\pi}}, &
  C_{110}^{1 -1} = -\displaystyle\frac{1}{8} \cdot \sqrt{\displaystyle\frac{3}{2\,\pi}}, &
  C_{110}^{11} = 0; \\

  \vspace{2mm}
  C_{111}^{-1 -1} = 0, &
  C_{111}^{-1 1} = -\displaystyle\frac{3}{8} \cdot \sqrt{\displaystyle\frac{1}{2\,\pi}}, &
  C_{111}^{1 -1} = \displaystyle\frac{3}{8} \cdot \sqrt{\displaystyle\frac{1}{2\,\pi}}, &
  C_{111}^{11} = 0; \\

  \vspace{2mm}
  C_{112}^{-1 -1} = \displaystyle\frac{1}{16} \: \sqrt{\displaystyle\frac{3}{2\,\pi}}, &
  C_{112}^{-1 1} = - \displaystyle\frac{1}{8} \sqrt{\displaystyle\frac{3}{2\,\pi}}, &
  C_{112}^{1 -1} = - \displaystyle\frac{1}{8} \sqrt{\displaystyle\frac{3}{2\,\pi}}, &
  C_{112}^{11} = \displaystyle\frac{1}{16} \sqrt{\displaystyle\frac{3}{2\,\pi}}.
\end{array}
\label{eq.app.4.7}
\end{equation}

\section{Functions $f_{l_{f}n}^{m \mu^{\prime}}(\theta)$
\label{app.5}}

\vspace{5mm}
Let us consider the function $f_{l_{f}n}^{m \mu^{\prime}}(\theta)$:
\begin{equation}
  f_{l_{f} n}^{m \mu^{\prime}}(\theta) =
    P_{l_{f}}^{1} (\cos{\theta}) \;  P_{1}^{1} (\cos{\theta}) \;  P_{n}^{|m+\mu^{\prime}|}(\cos{\theta}).
\label{eq.app.5.1}
\end{equation}
At $l_{f}=1$ and $n=0,1,2$ we obtain:
\begin{equation}
\begin{array}{lcl}
  f_{10}^{m \mu^{\prime}}(\theta) & = &
    P_{1}^{1} (\cos{\theta}) \; P_{1}^{1} (\cos{\theta}) \; P_{0}^{|m+\mu^{\prime}|} (\cos{\theta}), \\
  f_{11}^{m \mu^{\prime}}(\theta) & = &
    P_{1}^{1} (\cos{\theta}) \; P_{1}^{1} (\cos{\theta}) \; P_{1}^{|m+\mu^{\prime}|} (\cos{\theta}), \\
  f_{12}^{m \mu^{\prime}}(\theta) & = &
    P_{1}^{1} (\cos{\theta}) \; P_{1}^{1} (\cos{\theta}) \; P_{2}^{|m+\mu^{\prime}|} (\cos{\theta}).
\end{array}
\label{eq.app.5.2}
\end{equation}
At different $m = \pm 1$ and $\mu^{\prime} = \pm 1$ we find:
\begin{equation}
\begin{array}{lclcl}
  f_{10}^{-1, -1}(\theta) & = &
    P_{1}^{1} (\cos{\theta}) \;  P_{1}^{1} (\cos{\theta}) \;  P_{0}^{2} (\cos{\theta}) & = & 0, \\
  f_{10}^{-1 1}(\theta) & = &
    P_{1}^{1} (\cos{\theta}) \;  P_{1}^{1} (\cos{\theta}) \;  P_{0}^{0} (\cos{\theta}) & = &
    \sin^{2}{\theta}, \\
  f_{10}^{1 -1}(\theta) & = &
    P_{1}^{1} (\cos{\theta}) \;  P_{1}^{1} (\cos{\theta}) \;  P_{0}^{0} (\cos{\theta}) & = &
    \sin^{2}{\theta}, \\
  \vspace{3mm}
  f_{10}^{11}(\theta) & = &
    P_{1}^{1} (\cos{\theta}) \;  P_{1}^{1} (\cos{\theta}) \;  P_{0}^{2} (\cos{\theta}) & = & 0; \\

  f_{11}^{-1, -1}(\theta) & = &
    P_{1}^{1} (\cos{\theta}) \;  P_{1}^{1} (\cos{\theta}) \;  P_{1}^{2} (\cos{\theta}) & = & 0, \\
  f_{11}^{-1 1}(\theta) & = &
    P_{1}^{1} (\cos{\theta}) \;  P_{1}^{1} (\cos{\theta}) \;  P_{1}^{0} (\cos{\theta}) & = &
    \sin^{2}{\theta} \cos{\theta}, \\
  f_{11}^{1 -1}(\theta) & = &
    P_{1}^{1} (\cos{\theta}) \;  P_{1}^{1} (\cos{\theta}) \;  P_{1}^{0} (\cos{\theta}) & = &
    \sin^{2}{\theta} \cos{\theta}, \\
  \vspace{3mm}
  f_{11}^{1 1}(\theta) & = &
    P_{1}^{1} (\cos{\theta}) \;  P_{1}^{1} (\cos{\theta}) \;  P_{1}^{2} (\cos{\theta}) & = & 0; \\

  f_{12}^{-1, -1}(\theta) & = &
    P_{1}^{1} (\cos{\theta}) \;  P_{1}^{1} (\cos{\theta}) \;  P_{2}^{2} (\cos{\theta}) & = &
    3 \sin^{4}{\theta}, \\
  f_{12}^{-1 1}(\theta) & = &
    P_{1}^{1} (\cos{\theta}) \;  P_{1}^{1} (\cos{\theta}) \;  P_{2}^{0} (\cos{\theta}) & = &
    \frac{1}{2} \sin^{2}{\theta} \: (3\cos^{2}{\theta} - 1), \\
  f_{12}^{1 -1}(\theta) & = &
    P_{1}^{1} (\cos{\theta}) \;  P_{1}^{1} (\cos{\theta}) \;  P_{2}^{0} (\cos{\theta}) & = &
    \frac{1}{2} \sin^{2}{\theta} \: (3\cos^{2}{\theta} - 1), \\
  f_{12}^{11}(\theta) & = &
    P_{1}^{1} (\cos{\theta}) \;  P_{1}^{1} (\cos{\theta}) \;  P_{2}^{2} (\cos{\theta}) & = &
    3 \sin^{4}{\theta}.
\end{array}
\label{eq.app.5.3}
\end{equation}

\section{Dipole approach versus multipole one: comparative analysis and a role of the internal well of the potential
\label{app.6}}

After appearance of the fully quantum approach proposed by Papenbrock and Bertsch in~\cite{Papenbrock.1998.PRLTA} where wave function of photons was used in the dipole approximation, further fully quantum approaches have been developed mainly on its basis.
In particular, formula $\langle f|\mathbf{p}|i \rangle = i\hbar\, \langle f|\partial_{r} V|i \rangle /E_{\gamma}$ proposed in this paper for transformation of the matrix element of the photon emission and increasing essentially its convergence in calculations without visible decreasing of accuracy, becomes very popular.
This is essential point which attracts many researchers to study this problem in fully quantum approach.
Published results with agreement between spectra calculated by such dipole approach and experimental data look to be well.
In particular, consequence ``The present high precision data clearly demonstrate the failure of a classical Coulomb acceleration calculation (see, e.g., [9,14]) to describe the bremsstrahlung emission in $\alpha$ decay, and rules out theoretical suggestions put forward by authors of Refs.~[6,7,9]'' in~\cite{Boie.2007.PRL} (see p.~4, reference in cited paper) has been considered further as a fact that the dipole approach is the most accurate in description of experimental data and perspective from all fully quantum approaches in further study of bremsstrahlung emission during $\alpha$-decay.
However, let us clarify how much such approach is accurate in description of experimental data in comparison with the multipole one.

\subsection{Spectra in absolute scale
\label{app.6.1}}

\begin{figure}[htbp]
\centerline{%
\includegraphics[width=93mm]{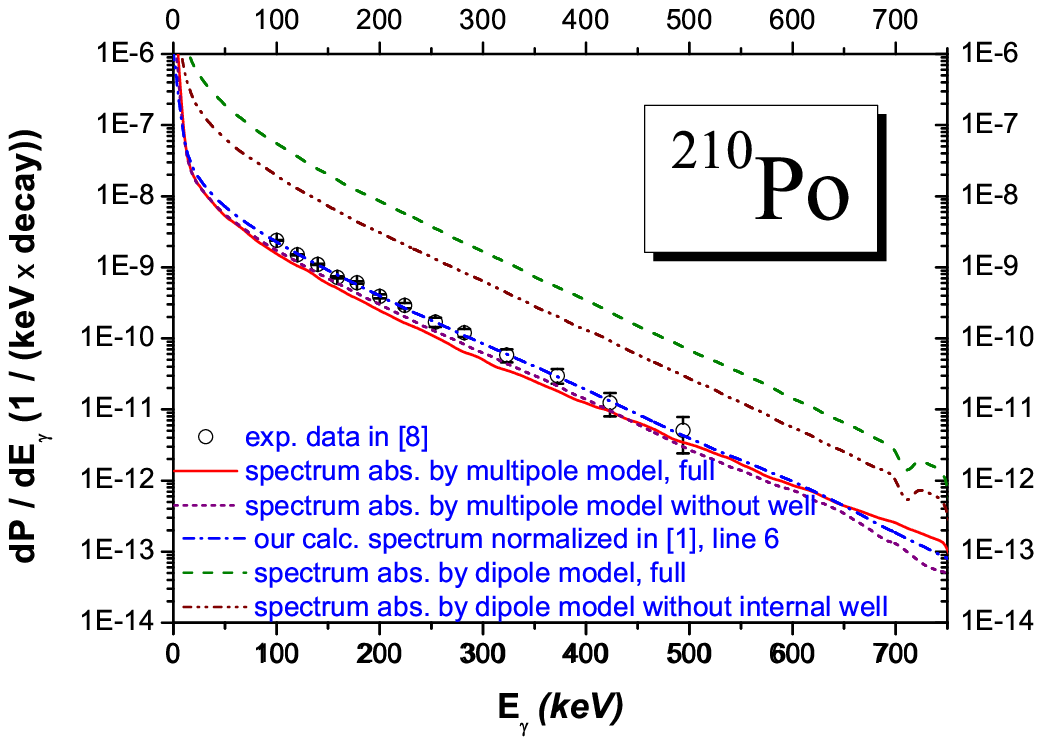}
\hspace{-3mm}\includegraphics[width=93mm]{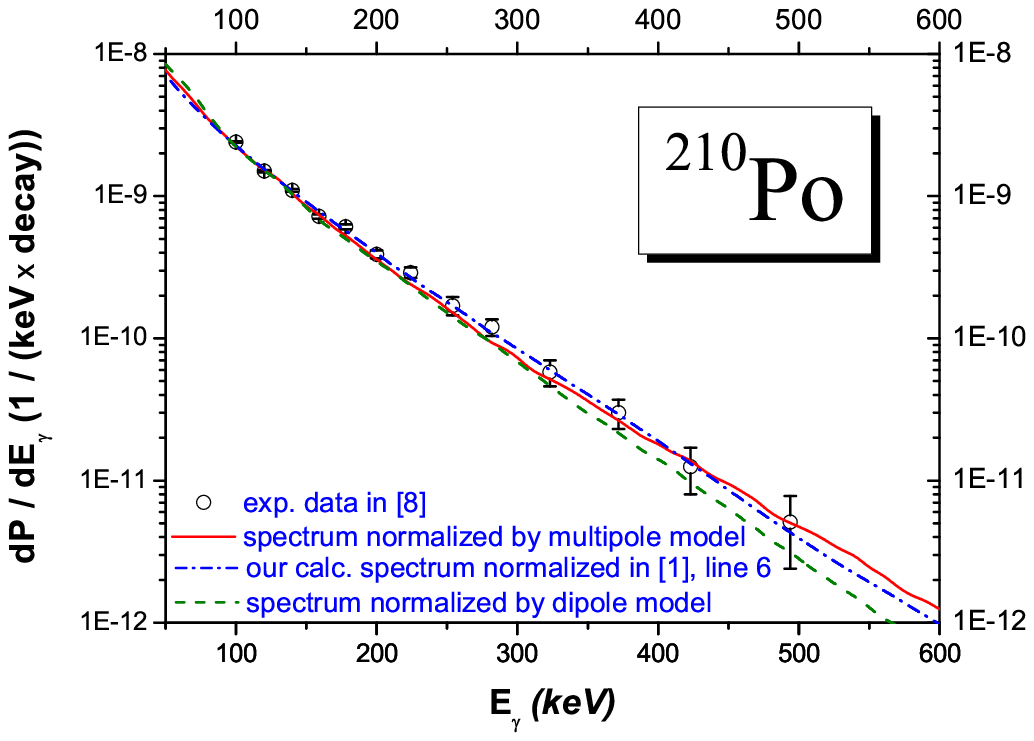}}
\vspace{-4mm}
\caption{Bremsstrahlung probability in the $\alpha$-decay of the $^{210}{\rm Po}$ nucleus
(solid line, red, is for the spectrum calculated by the multipole model;
dash line, green, for the full spectrum calculated by the dipole model;
dash-dot-dot line, brown, for the spectrum calculated by the dipole model without taking the internal region up to the internal turning point into account;
dash-dot line, blue, for the normalized spectrum calculated by the approach in \cite{Maydanyuk.2006.EPJA} with normalization used in this paper):
(left panel, a) Absolute probabilities: here one can see that difference between two spectra calculated in the dipole approach with inclusion of the internal region and without it is not small that confirms a real non-small influence of this nuclear region on the total spectrum;
(right panel, b) Normalized probabilities: here one can see that normalized spectra in multipole approach and in approach \cite{Maydanyuk.2006.EPJA} are found in better agreement with experimental data~\cite{Boie.2007.PRL} in comparison with normalized results obtained in the dipole approach (in contrast with conclusions of~\cite{Boie.2007.PRL}).
\label{fig.9}}
\end{figure}

Usually, authors do not mention whether they calculate normalized or absolute probability in the dipole approach. However, the spectra calculated in the absolute scale are sometimes noted to have a main progress of such a way. But if to suppose that these results were obtained without any normalization on experimental data, then I meet the following problem.
Up today, all published calculations in the dipole approach have been based on the $\alpha$-nucleus potential where rectangular well inside the internal region before the barrier was used. According to (\ref{eq.2.7.1.2}), we obtain directly null contribution of the emission from this internal region into the total spectrum. But such results would be reliable if the real emission from the internal region is very small. Only in such a case more accurate realistic shape of the well could be neglected.
But I find that this is not so.
Let us look on the left panel of Fig.~\ref{fig.9} where the spectra for the $^{210}{\rm Po}$ nucleus calculated by the dipole and multipole approaches without any normalization on experimental data are presented.
Here, one can see that difference between the full spectrum obtained by the dipole approach (see dash line, green, in this figure) and the spectrum obtained by the dipole approach formed in the spatial region starting from the internal turning point (see dash-dot-dot line, brown, in this figure) is not small (in calculations for $^{210}{\rm Po}$ such data are used: $Q_{\alpha} = 5.439$~MeV, internal turning point is $r_{\rm tp,\, int} = 8.829$~fm, external turning point is $r_{\rm tp,\, ext} = 44.007$~fm, tunneling length is $\Delta r_{\rm tun} = 35.177$~fm)! One can find that complete neglect of the emission from the internal region improves visibly agreement between the dipole calculations and experimental data. However, these two spectra are essentially farther to experimental data in comparison with the spectrum obtained by the multipole model without any normalization (see solid line, red, in this figure).
Now this confirms a real importance of inclusion of the realistic shape of the well before the barrier into calculations of the spectra in the absolute scale.
But if the spectra obtained previously in the dipole approach were in good or the best agreement with experimental data (authors are assumed to have catched a successful normalization in theory), then after taking the emission from the internal realistic well into account such spectra should be displaced below (to the distance similar to distance between two curves in the dipole approach in Fig.~\ref{fig.9}). As a result,
this demolishes all published progress in agreement between experimental data and calculations in the dipole approach. Now physical motivations of such progress in agreement between theory and experiment have became unclear if they were affirmed to be obtained in the absolute scale.
It has became clear which difficulty authors of \cite{Batkin.1986.SJNCA} had, who gave the first predictions for the $^{210}{\rm Po}$ nucleus before the first experiments~\cite{D'Arrigo.1994.PHLTA}.

\subsection{Spectra normalized on experimental data
\label{app.6.2}}

If to suppose that the spectra have obtained with normalization on experimental data, then we must compare them with results obtained in the multipolar approach with normalization also.
If in eq.~(\ref{eq.2.9.2}) for the absolute probability in the multipolar approach to restrict ourselves by the first most important integral $J(1,0)$ only in comparison with other two integrals $J(1,1)$ and $J(1,2)$, then we obtain (with possible normalization) the spectrum calculated by the model \cite{Maydanyuk.2006.EPJA} exactly.
Now exact coincidence has been established between the spectra obtained in the approach~\cite{Maydanyuk.2006.EPJA} and the spectra obtained by the multipolar model, where a difference is explained by taking non-zero magnetic component at $J(1,1)$ into account in the total matrix element in eqs.~(\ref{eq.2.9.1})
($J(1,2)$ is smaller then $J(1,0)$ by 1-2 orders from 50 keV up to 1 MeV of the photons).
The matrix element beyond the dipole approximation through expansion of the wave function of photons in \emph{spherical waves} (at first, proposed in~\cite{Maydanyuk.nucl-th.0404013}) and realistic form of the $\alpha$-nucleus potential realized in the model \cite{Maydanyuk.2006.EPJA} give us a small difference between the spectrum calculated for the $^{210}{\rm Po}$ nucleus by such a way and the spectrum obtained for this nucleus in the dipole approach with formula $\langle f|\mathbf{p}|i \rangle = i\hbar\, \langle f|\partial_{r} V|i \rangle /E_{\gamma}$ (see lines 6 and 7 in Fig.~1 in \cite{Maydanyuk.2006.EPJA} and comments here).
As latter experiment \cite{Boie.2007.PRL} for $^{210}{\rm Po}$ showed, this predicted peculiarity increases a little coincidence with experimental data of such paper in comparison with our calculations in the dipole approach above and with presented results (at lower energies of photons) in the fully quantum dipole approach in this paper also.
While authors of~\cite{Boie.2007.PRL} did not comment such comparison, did not present any grounds of their affirmation about groundlessness (ruling out) and accuracy of the approach~\cite{Maydanyuk.2006.EPJA},
the complete comparative consideration of these experimental data and all these spectra breaks down such affirmation (see Fig.~3 in \cite{Maydanyuk.2008.MPLA} and discussions). The same logics is applicable again for the multipole model directly. Let us look to the right panel in Fig.~\ref{fig.9}. Here, one can see that the spectrum calculated by the multipole model (see solid line, red, in this figure) looks to be a little more successful in description of the experimental data~\cite{Boie.2007.PRL} in comparison with the spectrum calculated by the dipole approach (see dash line, green, in this figure). Here, our discussed result by the approach~\cite{Maydanyuk.2008.MPLA} is included also (see dash-dot line, blue, in this figure). Results in descriptions of the newest experimental data~\cite{Maydanyuk.2008.EPJA,Maydanyuk.2008.MPLA,Maydanyuk.2009.NPA} for the $^{214}{\rm Po}$ and $^{226}{\rm Ra}$ nuclei in direction of the approach~\cite{Maydanyuk.2006.EPJA} look enough good also, where we have been achieving agreement between theory and experiment up to 765~keV (in \cite{Boie.2007.PRL} energy region of photons emitted is up to 500~keV).

Irrespective of accuracy which the approach could give to researcher, more important point consists in physics which such approach has.
While calculations of spectra in all other published approaches up today \cite{Takigawa.1999.PHRVA,Boie.2007.PRL,Jentschura.2008.PRC,Papenbrock.1998.PRLTA,Tkalya.1999.JETP,So_Kim.2000.JKPS} are based on the potential of interaction between the $\alpha$-particle and the daughter nucleus, which in the region of nuclear forces has a form of rectangular well and its depth is determined for each selected nucleus separately and absolutely does not take the real nuclear shape into account, our model and calculations starting from \cite{Maydanyuk.2006.EPJA} have been based on the unified global realistic nucleus--$\alpha$-particle potential which parameters are defined by $Q_{\alpha}$-value of the $\alpha$-decay, protons and neutrons numbers only for the studied nucleus. Such potential is constructed on the basis of analysis of 344 $\alpha$-decaying nuclei in frameworks of one unified model \emph{UMADAC} of the $\alpha$-decay and $\alpha$-capture \cite{Khudenko.2009.PRC.C79,Khudenko.2009.PRC.C80} (see also \cite{Denisov.2005.PHRVA}) where errors with experimental data turn out to be the smallest in comparison with other known models this year. This unification of the potential combines naturally with our formalism for calculations of the absolute angular (and integral) probabilities of bremsstrahlung during the $\alpha$-decay.
This allows to calculate the absolute bremsstrahlung probability for arbitrary nucleus using $Q_{\alpha}$-value, proton and neutron numbers only as input data.
After angular realization of the multipolar approach in calculation of the matrix elements, such a way allows to take a deformation of the $\alpha$-decaying nucleus into account \cite{Maydanyuk.2009.NPA}, which turns out to be \underline{not small} (that confirms a real importance of accurate determination of the $\alpha$-nucleus potential in the region of the nuclear forces) and it could be extracted from the bremsstrahlung spectra (what other published approaches in the dipole approach are not able to study in current their stage).
This fact underlines perspective of further investigations of the multipole approach. One can hope it should allow to open new independent experimental ways to obtain new information about $\alpha$-decay.


Argument presented above contradict to the affirmation of authors in~\cite{Boie.2007.PRL} about groundlessness of the approach~\cite{Maydanyuk.2006.EPJA} and results obtained by such approach. So, among all variety of existed models and approaches the multipolar model is the most motivated from the physical point of view, it is the richest in obtaining useful information about emission of photons during $\alpha$-decay, their results are in the best agreement with experimental data existed.


\section{Transformations of the matrix element in a general case
\label{app.7}}

The matrix element of emission (\ref{eq.2.7.1.1}) in the dipole approximation can be transformed as \cite{Papenbrock.1998.PRLTA}:
\begin{equation}
\begin{array}{lcl}
  \langle f\, |\, \mathbf{p}\, | i \rangle =
  -\,\displaystyle\frac
    {i\,\hbar\,
    \Bigl\langle f\, \Bigl|\,
    \displaystyle\frac{\partial\, U(\mathbf{r})}{\partial \mathbf{r}}\, \Bigr|\, i\, \Bigr\rangle }
    {E_{i} - E_{f}}.
\end{array}
\label{eq.app.7.1}
\end{equation}
In the resulting expression the integrant function obtains additional factor $1/r$ at enough far $r$ (in result of Coulomb type of the $\alpha$-nucleus potential here), that in the asymptotic region increases convergence of numerical integration of the total matrix element over $r$. So, it could be useful to generalize such a transformation for the multipole approach. However, in a general case the calculation of the matrix element looks to be more complicated and the following theorem can be applied.

\begin{Theorem}
Let us consider transition of the $\alpha$-decaying system from the initial $i$-state into the final $f$-state in result of emission of photon.
If the vector potential of photons was used in such a form
$\mathbf{A}\, (\mathbf{r}) =
\sqrt{\displaystyle\frac{2\pi \hbar c^{2}}{w}}\; \mathbf{e}^{(\alpha)}\, e^{i\,(\mathbf{kr}-wt)}$,
then the matrix element of such transition $i \to f$ would be written as
\begin{equation}
\begin{array}{lcl}
  \langle f\, |\, \mathbf{A}^{*}\, \mathbf{p}\, | i \rangle =
  -\,\displaystyle\frac
    {i\,\hbar\,
    \Bigl\langle f\, \Bigl|\,
      \mathbf{A}^{*}\,
      \displaystyle\frac{\partial\, U(\mathbf{r})}{\partial \mathbf{r}}\, \Bigr|\, i\, \Bigr\rangle +
      \displaystyle\frac{\hbar^{2}}{m}\:
      \Bigl\langle\, f\, \Bigl|\,
      \biggl(\mathbf{A}^{*}\, \displaystyle\frac{\partial}{\partial \mathbf{r}} \biggr)\,
      \biggl(\mathbf{k} \displaystyle\frac{\partial}{\partial \mathbf{r}} \biggr) \Bigr|\, i\,
      \Bigr \rangle}
    {E_{i} - E_{f} + \displaystyle\frac{\hbar\, k^{2}}{2m}},
\end{array}
\label{eq.app.7.2}
\end{equation}
where $U\,(\mathbf{r})$ is the $\alpha$-nucleus potential,
$m$ is reduced mass,
$E_{i}$ and $E_{f}$ are energies of the $\alpha$-decaying system in the initial $i$-state and in the final $f$-state, $\mathbf{k}$ is wave vector of photon emitted, $k = |\mathbf{k}|$ is its wave number.
\end{Theorem}

\begin{proof}
Let us consider the following commutator:
\begin{equation}
\begin{array}{lcl}
\vspace{2mm}
  [\hat{H},\, \mathbf{A}^{*}]\: \varphi(\mathbf{r}) & = &
  \hat{H} \mathbf{A}^{*}\, \varphi(\mathbf{r}) - \mathbf{A}^{*} \hat{H}\, \varphi(\mathbf{r}) =
    \Bigl( \displaystyle\frac{\mathbf{\hat{p}}^{2}}{2m} + U(\mathbf{r}) \Bigr) \cdot \mathbf{A}^{*}
    \varphi(\mathbf{r}) -
    \mathbf{A}^{*} \cdot
    \Bigl( \displaystyle\frac{\mathbf{\hat{p}}^{2}}{2m} + U(\mathbf{r}) \Bigr) \cdot \varphi(\mathbf{r}) = \\

\vspace{2mm}
  & = &
    \Bigl( - \displaystyle\frac{\hbar}{2m} \displaystyle\frac{\partial^{2}}{\partial \mathbf{r}^{2}} \Bigr)
    \cdot \mathbf{A}^{*}(\mathbf{r})\, \varphi(\mathbf{r}) -
    \mathbf{A}^{*}(\mathbf{r}) \cdot
    \Bigl( - \displaystyle\frac{\hbar}{2m} \displaystyle\frac{\partial^{2}}{\partial \mathbf{r}^{2}} \Bigr)
    \cdot \varphi(\mathbf{r}) = \\

\vspace{2mm}
  & = &
    - \displaystyle\frac{\hbar}{2m}
    \biggl( \displaystyle\frac{\partial^{2}}{\partial \mathbf{r}^{2}}\: \mathbf{A}^{*}(\mathbf{r}) \biggr)\,
    \varphi(\mathbf{r}) -
    2\, \displaystyle\frac{\hbar}{2m}
    \biggl( \displaystyle\frac{\partial}{\partial \mathbf{r}}\: \mathbf{A}^{*}(\mathbf{r}) \biggr)
    \biggl( \displaystyle\frac{\partial}{\partial \mathbf{r}}\: \varphi(\mathbf{r}) \biggr).
\end{array}
\label{eq.app.7.3}
\end{equation}
Taking explicit form of $\mathbf{A}\,(\mathbf{r})$ into account, we find
\begin{equation}
\begin{array}{lcl}
\vspace{2mm}
  \biggl( \displaystyle\frac{\partial^{2}}{\partial \mathbf{r}^{2}}\, \mathbf{A}^{*}(\mathbf{r}) \biggr)\,
  \varphi(\mathbf{r}) & = &
  \biggl(
    \displaystyle\frac{\partial}{\partial \mathbf{r}}
    \displaystyle\frac{\partial}{\partial \mathbf{r}}
    \sqrt{\displaystyle\frac{2\pi \hbar c^{2}}{w}}\, \mathbf{e}^{(\alpha),*} e^{-i\,(\mathbf{kr}-wt)}
  \biggr)
  \varphi(\mathbf{r}) =

  \biggl(
    (-i \mathbf{k})
    \displaystyle\frac{\partial}{\partial \mathbf{r}}
    \sqrt{\displaystyle\frac{2\pi \hbar c^{2}}{w}}\: \mathbf{e}^{(\alpha),*} e^{-i\,(\mathbf{kr}-wt)}
  \biggr)
  \varphi(\mathbf{r}) = \\

\vspace{2mm}
  & = &
  \biggl(
    (- \mathbf{k}^{2}) \sqrt{\displaystyle\frac{2\pi \hbar c^{2}}{w}}\: \mathbf{e}^{(\alpha),*} e^{-i\,(\mathbf{kr}-wt)}
  \biggr)
  \varphi(\mathbf{r}) =
  - k^{2}\, \mathbf{A}^{*} \, \varphi(\mathbf{r}), \\

\vspace{2mm}
  \biggl( \displaystyle\frac{\partial}{\partial \mathbf{r}}\, \mathbf{A}^{*}(\mathbf{r}) \biggr)
    \biggl( \displaystyle\frac{\partial}{\partial \mathbf{r}}\, \varphi(\mathbf{r}) \biggr) & = &
  \biggl( \displaystyle\frac{\partial}{\partial \mathbf{r}}
    \sqrt{\displaystyle\frac{2\pi \hbar c^{2}}{w}}\, \mathbf{e}^{(\alpha),*} e^{-i\,(\mathbf{kr}-wt)}
  \biggr)
  \biggl( \displaystyle\frac{\partial}{\partial \mathbf{r}}\, \varphi(\mathbf{r}) \biggr) = \\

  & = &
  \biggl( -i \mathbf{k}
    \sqrt{\displaystyle\frac{2\pi \hbar c^{2}}{w}}\, \mathbf{e}^{(\alpha),*} e^{-i\,(\mathbf{kr}-wt)}
  \biggr)
  \biggl( \displaystyle\frac{\partial}{\partial \mathbf{r}}\, \varphi(\mathbf{r}) \biggr) =

  -i\,\mathbf{A}^{*}
  \biggl(\mathbf{k} \displaystyle\frac{\partial}{\partial \mathbf{r}} \biggr)\varphi(\mathbf{r})
\end{array}
\label{eq.app.7.4}
\end{equation}
and from (\ref{eq.app.7.3}) we obtain
\[
\begin{array}{lcl}
  \hat{H} \mathbf{A}^{*}\, \varphi(\mathbf{r}) - \mathbf{A}^{*} \hat{H}\, \varphi(\mathbf{r}) & = &
    \displaystyle\frac{\hbar\, k^{2}}{2m}\: \mathbf{A}^{*}(\mathbf{r})\, \varphi(\mathbf{r}) +
    \displaystyle\frac{i\,\hbar}{m}\, \mathbf{A}^{*}\,
      \biggl(\mathbf{k} \displaystyle\frac{\partial}{\partial \mathbf{r}}\biggr)\, \varphi(\mathbf{r})
\end{array}
\]
or
\begin{equation}
\begin{array}{lcl}
  \hat{H} \mathbf{A}^{*}\, \varphi(\mathbf{r}) =
  \biggl\{
    \mathbf{A}^{*} \hat{H} +
    \displaystyle\frac{\hbar\, k^{2}}{2m}\: \mathbf{A}^{*}(\mathbf{r}) +
    \displaystyle\frac{i\,\hbar}{m}\, \mathbf{A}^{*}\,
    \biggl(\mathbf{k} \displaystyle\frac{\partial}{\partial \mathbf{r}}\biggr)
  \biggr\}\,
  \varphi(\mathbf{r}).
\end{array}
\label{eq.app.7.5}
\end{equation}
Now we write ($\mathbf{p}\, \mathbf{A}^{*} = \mathbf{A}^{*} \mathbf{p} \ne 0$ in Coulomb gauge)
\begin{equation}
\begin{array}{lcl}
\vspace{2mm}
  \langle f\, |\, \mathbf{p}\, \hat{H}\, \mathbf{A}^{*}\, | i \rangle & = &
  \langle f\, |\, \mathbf{p}\, \mathbf{A}^{*}\, \hat{H}\, | i \rangle +
    \displaystyle\frac{\hbar\, k^{2}}{2m}\: \langle f\, |\, \mathbf{p}\, \mathbf{A}^{*}\, | i \rangle +
    \displaystyle\frac{i\,\hbar}{m}\,
    \langle f\, |\, \Bigl(\mathbf{p}\, \mathbf{A}^{*}\Bigr)\,
      \biggl(\mathbf{k} \displaystyle\frac{\partial}{\partial \mathbf{r}} \biggr) \, | i \rangle = \\
\vspace{2mm}
  & = &
    \langle f\, |\, \mathbf{p}\, \mathbf{A}^{*}\, E_{i}\, | i \rangle +
    \displaystyle\frac{\hbar\, k^{2}}{2m}\: \langle f\, |\, \mathbf{p}\, \mathbf{A}^{*}\, | i \rangle +
    \displaystyle\frac{i\,\hbar}{m}\,
    \langle f\, |\, \Bigl(\mathbf{p}\, \mathbf{A}^{*}\Bigr)\,
      \biggl(\mathbf{k} \displaystyle\frac{\partial}{\partial \mathbf{r}} \biggr) \, | i \rangle = \\
  & = &
    \biggl(E_{i} + \displaystyle\frac{\hbar\, k^{2}}{2m} \biggr) \cdot
    \langle f\, |\, \mathbf{A}^{*}\, \mathbf{p}\, | i \rangle +
    \displaystyle\frac{\hbar^{2}}{m} \cdot
    \Bigl\langle\, f\, \Bigl|\,
      \biggl(\mathbf{A}^{*}\, \displaystyle\frac{\partial}{\partial \mathbf{r}} \biggr)\,
      \biggl(\mathbf{k} \displaystyle\frac{\partial}{\partial \mathbf{r}} \biggr) \Bigr|\, i\,
    \Bigr\rangle
\end{array}
\label{eq.app.7.6}
\end{equation}
and we have also
\begin{equation}
\begin{array}{lcl}
\vspace{2mm}
  \langle f\, |\, \hat{H}\, \mathbf{p}\, \mathbf{A}^{*}\, | i \rangle =
  \langle f\, |\, E_{f}\, \mathbf{p}\, \mathbf{A}^{*}\, | i \rangle =
  E_{f} \cdot \langle f |\, \mathbf{p}\, \mathbf{A}^{*}\, | i \rangle.
\end{array}
\label{eq.app.7.7}
\end{equation}
From (\ref{eq.app.7.6}) and (\ref{eq.app.7.7}) we obtain:
\begin{equation}
\begin{array}{lcl}
\vspace{2mm}
  \langle f\, |\, [\hat{H},\, \mathbf{p}]\, \mathbf{A}^{*}\, | i \rangle =

  \langle f\, |\, \hat{H}\, \mathbf{p}\, \mathbf{A}^{*}\, | i \rangle -
  \langle f\, |\, \mathbf{p}\, \hat{H}\, \mathbf{A}^{*}\, | i \rangle = \\

  = E_{f} \cdot \langle f |\, \mathbf{p}\, \mathbf{A}^{*}\, | i \rangle -
    \biggl(E_{i} + \displaystyle\frac{\hbar\, k^{2}}{2m} \biggr) \cdot
    \langle f\, |\, \mathbf{A}^{*}\, \mathbf{p}\, | i \rangle -
    \displaystyle\frac{\hbar^{2}}{m} \cdot
    \Bigl\langle\, f\, \Bigl|\,
      \biggl(\mathbf{A}^{*}\, \displaystyle\frac{\partial}{\partial \mathbf{r}} \biggr)\,
      \biggl(\mathbf{k} \displaystyle\frac{\partial}{\partial \mathbf{r}} \biggr) \Bigr|\, i\,
    \Bigr\rangle = \\

  = - \biggl(E_{i} - E_{f} + \displaystyle\frac{\hbar\, k^{2}}{2m} \biggr) \cdot
    \langle f\, |\, \mathbf{A}^{*}\, \mathbf{p}\, | i \rangle -
    \displaystyle\frac{\hbar^{2}}{m} \cdot
    \Bigl\langle\, f\, \Bigl|\,
      \biggl(\mathbf{A}^{*}\, \displaystyle\frac{\partial}{\partial \mathbf{r}} \biggr)\,
      \biggl(\mathbf{k} \displaystyle\frac{\partial}{\partial \mathbf{r}} \biggr) \Bigr|\, i\,
    \Bigr\rangle
\end{array}
\label{eq.app.7.8}
\end{equation}
and from here we find:
\begin{equation}
\begin{array}{lcl}
  \langle f\, |\, \mathbf{A}^{*}\, \mathbf{p}\, | i \rangle =
  -\,\displaystyle\frac
    {\langle f\, |\, [\hat{H},\, \mathbf{p}]\, \mathbf{A}^{*}\, | i \rangle +
      \displaystyle\frac{\hbar^{2}}{m} \cdot
      \Bigl\langle\, f\, \Bigl|\,
      \biggl(\mathbf{A}^{*}\, \displaystyle\frac{\partial}{\partial \mathbf{r}} \biggr)\,
      \biggl(\mathbf{k} \displaystyle\frac{\partial}{\partial \mathbf{r}} \biggr) \Bigr|\, i\,
      \Bigr \rangle}
    {E_{i} - E_{f} + \displaystyle\frac{\hbar\, k^{2}}{2m}}.
\end{array}
\label{eq.app.7.9}
\end{equation}
Now taking into account
\begin{equation}
  [\hat{H},\, \mathbf{p}] = i\,\hbar\, \displaystyle\frac{\partial\, U(\mathbf{r})}{\partial \mathbf{r}},
\label{eq.app.7.10}
\end{equation}
we obtain (\ref{eq.app.7.2}).
%
\end{proof}
Physical sense of the formula (\ref{eq.app.7.2}) could be explained by the following. The first item in this formula seems to have higher convergence  in the asymptotic region at numerical integration over $r$. It is supposed to give major contribution into the total matrix element while the second item allows us to analyze corrections after taking more accurate estimation of the emission of photons in the far asymptotic region into account (which is smaller usually).




\end{document}